\documentclass[prb,twocolumn,groupedaddress,aps,amsmath,amsfonts,notitlepage,longbibliography]{revtex4-1}
\usepackage{amsmath,amsfonts,amssymb,dsfont,graphicx,bm}
\graphicspath{}
\usepackage{color}
\usepackage{hyperref}
\hypersetup{
    bookmarks=false,         % show bookmarks bar?
    unicode=false,          % non-Latin characters in AcrobatÕs bookmarks
    pdftoolbar=false,        % show AcrobatÕs toolbar?
    pdfmenubar=true,        % show AcrobatÕs menu?
    pdffitwindow=false,     % window fit to page when opened
    pdfstartview={FitH},    % fits the width of the page to the window
    pdftitle={Gully quantum Hall ferromagnetism in biased trilayer graphene},    % title
    pdfauthor={P. Rao and M. Serbyn},     % author
    pdfsubject={},   % subject of the document
    pdfcreator={},   % creator of the document
    pdfproducer={}, % producer of the document
    pdfkeywords={quantum Hall} {symmetry breaking}{graphene}, % list of keywords
    pdfnewwindow=true,      % links in new window
    colorlinks=true,       % false: boxed links; true: colored links
    linkcolor=black,          % color of internal links (change box color with linkbordercolor)
    citecolor=blue,        % color of links to bibliography
    filecolor=magenta,      % color of file links
    urlcolor=blue           % color of external links
}

\def\be{\begin{equation}}
\def\ee{\end{equation}}
\def\bea{\begin{eqnarray}}
\def\eea{\end{eqnarray}}
\def\bpm{\begin{pmatrix}}
\def\epm{\end{pmatrix}}
\def\diag{\mathop{\rm diag}}

\newcommand{\ket}[1]{\left| #1\right\rangle}

\begin{document}
\title{Gully quantum Hall ferromagnetism in biased trilayer graphene}
\author{Peng Rao}
\author{Maksym Serbyn} 
\affiliation{Institute of Science and Technology Austria, 3400 Klosterneuburg, Austria}
\date{\today}
\begin{abstract}
Multilayer graphene lattices allow for an additional tunability of the band structure by the strong perpendicular electric field. In particular, the emergence of the new multiple Dirac points in ABA stacked trilayer graphene subject to strong transverse electric fields was proposed theoretically and confirmed experimentally. These new Dirac points dubbed ``gullies'' emerge from the interplay between strong electric field and trigonal warping. In this work we first characterize the properties of new emergent Dirac points and show that the electric field can be used to tune the distance between gullies in the momentum space. We demonstrate that the band structure has multiple Lifshitz transitions and higher-order singularity of ``monkey saddle'' type. Following the characterization of the band structure, we consider the spectrum of Landau levels and structure of their wave functions. In the limit of strong electric fields when gullies are well separated in momentum space, they give rise to triply degenerate Landau levels. In the second part of this work, we investigate how degeneracy between three gully Landau levels is lifted in presence of interactions. Within the Hartree-Fock approximation we show that the symmetry breaking state interpolates between fully gully polarized state that breaks $C_3$ symmetry at high displacement field, and the gully symmetric state when the electric field is decreased. The discontinuous transition between these two states is driven by enhanced inter-gully tunneling and exchange. We conclude by outlining specific experimental predictions for the existence of such a symmetry-breaking state. 
\end{abstract}

\maketitle

\section{Introduction \label{Sec:Intro}}

Since experimental realization of graphene,~\cite{Novoselov2004} two dimensional materials have been a focus of intense research. The single-layer graphene band structure provided realization of four copies of Dirac fermions. Moving from single layer graphene to multilayer graphene lattices, it was demonstrated that one can realize massive Dirac fermions,~\cite{McCann2006} Dirac fermions with (approximately) cubic dispersion~\cite{McCann2009,Fan2010} and combination of massive and massless Dirac fermions.~\cite{McCannGateABA}   Additional tunability of the band structure can be achieved by applying transverse electric field. For the bilayer graphene it leads to the gap opening.~\cite{Castro2007} For stronger electric fields, the interplay between the field and trigonal warping was predicted to lead to the new set of emergent Dirac points in both bilayer~\cite{McCann2006} and ABA-stacked trilayer graphene.~\cite{Maksym2013,morimoto}

Recently the emergence of new Dirac points was demonstrated experimentally for the ABA-stacking trilayer graphene (TLG).~\cite{Rao2018} Under strong external electric field, the low-energy band structure consists of multiple band minima or ``gullies'' (maxima for hole-like bands) that come in triples due to $C_3$ rotational symmetry. Moreover, the position of these gullies in the momentum space is tunable by the strength of electric field. In a presence of sufficiently weak perpendicular magnetic field, such gullies would lead to 3-fold degenerate Landau levels. 

Similar gully configurations have also been reported in a number of systems, i.e. SnTe-(111),~\cite{Li2016} PbTe-(111),~\cite{Chitta2006} and Bi-(111) surfaces.~\cite{Koroteev2004} Presence of interactions is expected to split this degeneracy giving symmetry-broken states. Ref.~\onlinecite{Sodemann2017} suggested that these symmetry broken states must be maximally ``gully polarized'', e.g.~that they are completely concentrated in one gully if distance between gullies in momentum space is sufficiently large, so that one can neglect inter-gully electron scattering.  However, this condition is not satisfied for the case of ABA graphene in the case of weak electric field or strong magnetic fields. 

In this work we consider the interaction effects on the gully degenerate Landau levels in the ABA trilayer graphene. To this end we begin with characterization of the non-interacting band structure of ABA trilayer graphene in presence of strong electric field. The evolution of the band structure upon application of strong transverse electric field was considered in Refs.~\onlinecite{Maksym2013,morimoto}. However, Ref.~\onlinecite{Maksym2013} focused mostly on the regime of strong magnetic fields, whereas Ref.~\onlinecite{morimoto} concentrated on study of the valley Hall state and associated edge states. In contrast, here we focus on understanding different parameters of the new emergent gullies that are relevant for the interaction effects. We explore dependence of gully parameters on the strength of electric field. In addition, we illustrate the presence of multiple Lifshitz transitions in the band structure and also emergence of higher-order singularity when three van Hove singularities meet with each other. 

After characterization of band structure, we discuss the spectrum of Landau levels. We focus on the three-fold degenerate (provided one ignores spin) Landau levels in the regime of weak magnetic fields. Although these LLs were observed before,~\onlinecite{Maksym2013,morimoto} we investigate their structure in greater detail. In particular, we discuss the splitting of LLs due to magnetic breakdown and also study the form of individual Landau levels wave functions since it controls the interaction effects via form-factors.~\cite{MacDonald1986} 

After providing basic understanding of the band structure and Landau level spectrum, we consider the effects of interactions on the three-fold degenerate Landau level at filling $\nu=1$.~(We assume full spin polarization which is favoured by exchange and Zeeman energies.) Analytically we find that the ground state at this filling factor is either polarized in one gully, thus breaking $C_3$ symmetry, or is a coherent $C_3$ symmetric superposition of states in all three gullies. The inter-gully scattering as well as tunneling between gullies, which cannot be neglected for small inter-gully distances lead to violation of the ``gully polarization theorem'' proposed in Ref.~\onlinecite{Sodemann2017}. We set up a self-consistent Hartree-Fock scheme that takes into account both inter-gully scattering and tunneling effects. Our calculations show that HF ground state undergoes the first order phase  transition as a function of electric field. Thus we conclude that the ABA-stacked trilayer graphene provides a perspective platform for probing the first order nematic transition where spontaneous (partial) gully polarization develops. 

Our work is inspired by the experiment~\cite{Rao2018} that confirmed presence of emergent Dirac gullies and suggested the presence of symmetry broken states.  
We predict that these states can be characterized by a non-vanishing expectation value of the dipole moment. Motivated by the experimental setup that includes encapsulated graphene, we consider the limit of screened Coulomb interaction, where it plays a subleading role compared to single-particle splittings. We note, that recent work also investigated the qualitatively different regime of strong interactions in the suspended multilayer graphene samples,\cite{Nam2018} where interactions lead to gap opening even without magnetic field.

The remainder of the paper is arranged as follows. In Sec.~\ref{Sec:BandStructure}, we introduce the tight-binding model, discuss the band structure and Fermi-surface topology in the absence of magnetic field. Section~\ref{Sec:MagneticField} considers behavior of the Landau level spectrum in weak and strong magnetic field limits. We show that at large external electric field, triply-degenerate Landau levels (for one spin component) are formed corresponding to sets of gullies related by $C_3$ symmetry. In addition we discuss the structure of the Landau level wave functions, since it is important for determining the interaction effects. Finally, Sec.~\ref{Sec:InteractionEffects} considers interaction effects within the Hartree-Fock approximation. We present analytical calculations using simple model in the gully basis. Later these calculations are compared with numerical results from self-consistent Hartree-Fock approximation. We conclude in Sec.~\ref{Sec:Discussion} by discussing experimental implications of our results.

\section{Review of band structure and emergent Dirac gullies \label{Sec:BandStructure}}
In this section we discuss the properties of new emergent Dirac points that were predicted in Refs.~\onlinecite{Maksym2013,morimoto} in the non-interacting band structure of ABA graphene subject to a strong electric field. We concentrate on their physical properties, i.e.\ anisotropy, using tight-binding parameters from recent experiments. The new set of tight binding parameters used here results in predictions  that differ from earlier studies.~\cite{Maksym2013,morimoto} In addition, we also discuss Lifshitz transitions and report existence of the higher order singularity that was previously theoretically studied in bilayer graphene.~\cite{Shtyk2017}
\subsection{Tight-binding model and band structure}
We use the Slonczewski-Weiss-McClure parametrization of the tight-binding model introduced in Ref.~\onlinecite{Dress2002} to describe the band structure of ABA trilayer graphene. The tight binding description requires a six-atom basis corresponding to 2 sublattices in three different layers. Via a suitable rotation of the basis,  the $6\times6$ tight-binding Hamiltonian can be brought to the block form consisting of single-layer (SLG) and bilayer graphene (BLG) like blocks, mixed by the external electric field $\Delta_1$:
\begin{equation}
\label{Eq:Ham}
H = \begin{pmatrix}H_{\text{SLG}} & V_{\Delta_1} \\ V^T_{\Delta_1} & H_{\text{BLG}}\end{pmatrix}.
\end{equation}
Hamiltonians of respective blocks read: 
\begin{equation}\label{Hamiltonian1}
\begin{aligned}
&H_\text{SLG}= 
\begin{pmatrix} \Delta_2 - \frac{\gamma_2}{2} & v_0 \pi^{\dagger} \\
v_0 \pi & -\frac{\gamma_5}{2}+\delta + \Delta_2 \end{pmatrix}, \\
&H_\text{BLG} =\\
&\begin{pmatrix} \frac{\gamma_2}{2}+ \Delta_2 & \sqrt{2} v_3 \pi & -\sqrt{2} v_4 \pi^{\dagger} & v_0 \pi^{\dagger}\\ \sqrt{2} v_3 \pi^{\dagger} & -2\Delta_2 &  v_0 \pi & -\sqrt{2} v_4 \pi \\
-\sqrt{2} v_4 \pi & v_0 \pi^{\dagger} & \delta-2\Delta_2& \sqrt{2} \gamma_1 \\
v_0\pi & -\sqrt{2} v_4 \pi^{\dagger}& \sqrt{2}\gamma_1 & \frac{\gamma_5}{2} + \delta + \Delta_2
\end{pmatrix},
\end{aligned}
\end{equation}
where tight-binding parameters $v_0, \gamma_1, \gamma_2, v_3, v_4, \gamma_5, \delta, \Delta_2$ and the momentum-dependent function $\pi$ are described in the Appendix~\ref{Appendix:Bandstructure}. In this work we use the value of these parameters from Ref.~\onlinecite{Rao2018} where they were determined by fits to experimental data.

The matrix that is responsible for mixing between SLG and BLG blocks is proportional to the potential difference induced by transverse electric field, $\Delta_1$,
\begin{equation}
V_{\Delta_1}  = \begin{pmatrix}
\Delta_1&0&0&0 \\ 0&0&0&\Delta_1
\end{pmatrix}.
\end{equation}
In the absence of transverse  electric field $\Delta_1=0$ and  the SLG and BLG blocks are independent and resulting low-energy band structure consists of SLG-like linearly dispersing band and BLG-like quadratically dispersing band.~\cite{McCannGateABA,Maksym2013,morimoto} We note both of that these low energy bands are generally gapped and displaced with respect to each other. 

\begin{figure*}
	\includegraphics*[width=\linewidth]{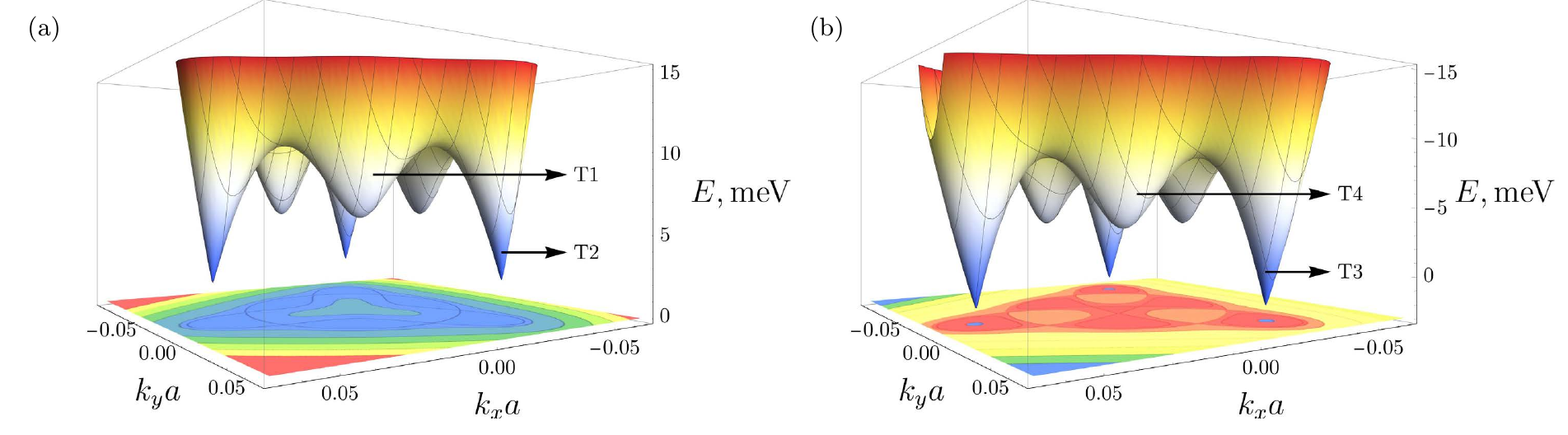}
    \caption{\label{Fig:Bandstructure} Three dimensional plot of electron (a) and hole (b) band structure plot at $\Delta_1=100$~meV with their projected contour plots near $K^+$ point. Energy axis is inverted in the hole band (b) for convenience. The gullies are labeled in order of decreasing energy as T1-T4.
    }   
\end{figure*}

\subsection{Band structure in the gully-limit}
When the ABA graphene sheet is subject to the perpendicular electric field, the non-zero matrix $V_{\Delta_1}$ in Eq.~(\ref{Eq:Ham}) hybridizes the SLG and BLG bands. In addition, the SLG-like band rapidly floats away from the neutrality point as $\Delta_1$ increases. In the limit of sufficiently high $\Delta_1\geq 30 $~meV~(corresponding to electric field strength $\sim0.15$~V/nm),  
the interplay of trigonal warping and electric field gives rise to a set of new emergent Dirac points that we dub ``gullies'' in what follows. These gullies have been discovered in Refs.~\onlinecite{Maksym2013,morimoto}. Here we concentrate on their physical properties, i.e. anisotropy, using tight-binding parameters from recent experiments.

In Refs.~\onlinecite{Maksym2013,morimoto}, it was demonstrated that these emergent gullies at large $\Delta_1$ can be understood from the so-called chiral limit. In this limit, one retains only large tight-binding parameters $v_0,\gamma_1, v_3, \Delta_1$ leading to the particle-hole symmetric band structure. Then the original Dirac points at $K^\pm$ valleys split into six off-centered massless Dirac points and a central Dirac point. By including the previously neglected tight-binding parameters, one breaks the particle-hole symmetry, making the electron and hole band structures different from each other. In addition, the  tight-binding parameters that were neglected in the chiral approximation, break the symmetry between six off-center Dirac points splitting them into two different sets each containing three Dirac points [see Fig.~\ref{Fig:Bandstructure}]. The three Dirac points within each set are related by $C_3$ rotation symmetry. These two sets of off-center Dirac points differ from each other by values of gap and other parameters, as will be discussed below. 

\begin{figure}[b]
\includegraphics*[width=0.95\linewidth]{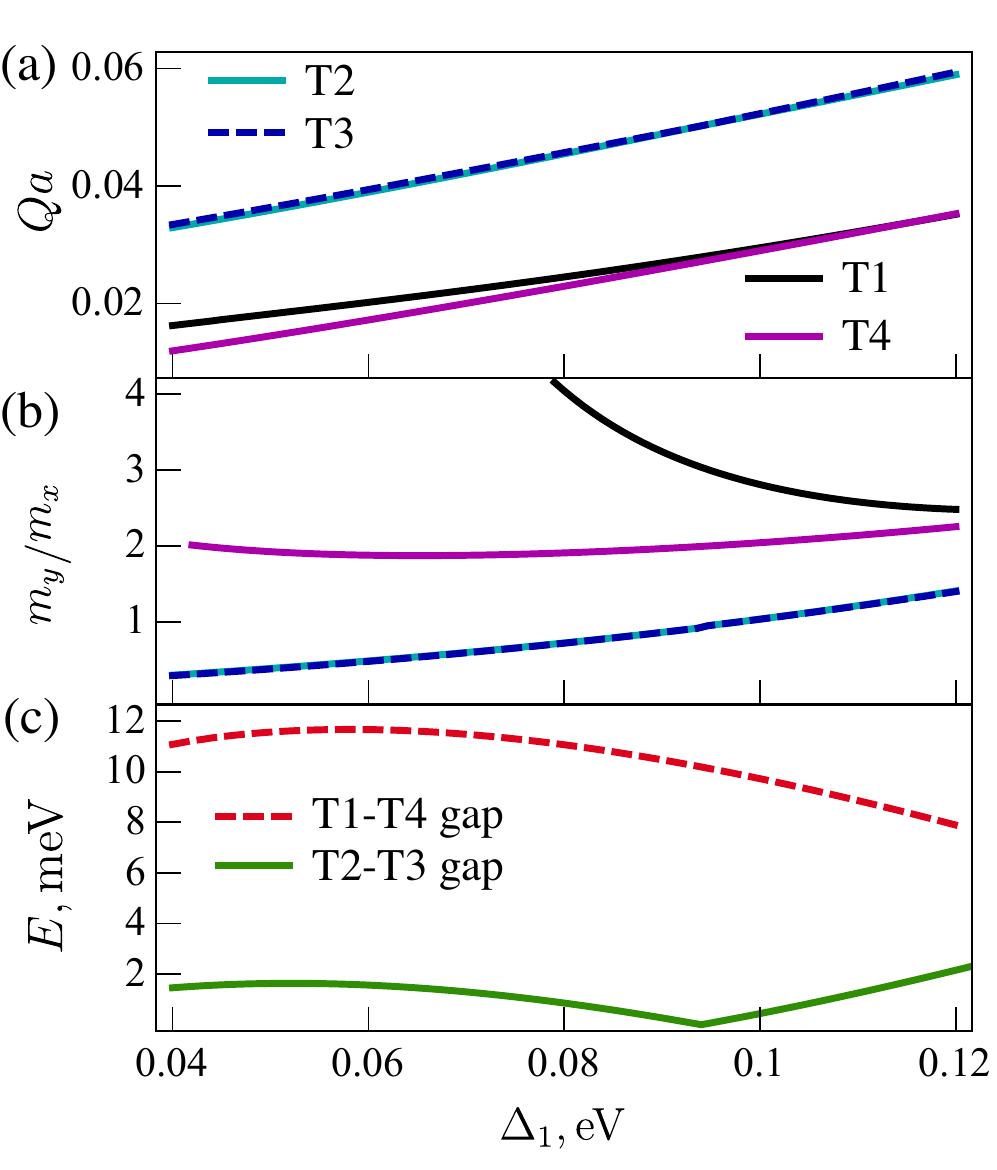}
	\caption{\label{Fig:Gullies}   (a) The displacement of the gullies center relative to the $K$ point, $Qa$  is monotonously increasing with electric field $\Delta_1$.
	(b) Effective mass ratios reveal very anisotropic character of T1 gully in contrast to its hole counterpart T4. (c) Gap between T2-T3 and T1-T4 gullies has non-monotonous dependence with electric field. The gap closure happen for $\Delta_1=92$~meV for gullies T2-T3 and at $\Delta_1=185$~meV between T1-T4 which is not shown here.
	}    
\end{figure}

We label these gullies in order of decreasing energy as T1-T4, see Fig.~\ref{Fig:Bandstructure}. On the electron side in Fig.~\ref{Fig:Bandstructure}(a), the inner gully would be T$1$ while the outer one T$2$. On the hole side in Fig.~\ref{Fig:Bandstructure}(b), the inner gully is T$4$ and outer one T$3$. The gullies' positions and anisotropies are characterized respectively by their distance to the $K^\pm$ points, the gap between each two approximately particle-hole symmetric sets (T$1$, T$4$ and T$2$, T$3$) and their effective mass ratios. These three parameters are plotted as functions of $\Delta_1$ in Fig.~\ref{Fig:Gullies}. While the field dependence of Dirac mass was considered before,\cite{Maksym2013,morimoto} the anisotropy of effective masses and distance of the Dirac points from the $K$ point in the reciprocal state were not investigated. Moreover, due to the different set of tight-binding parameters, the gap closure between T2 and T3 gullies happens at electric fields that at least factor two smaller compared to previous estimates.\cite{Maksym2013,morimoto} The second set of gullies is also predicted to have a gap closure at higher values of electric fields.
  
\subsection{Lifshitz Transitions and Monkey Saddle}
\begin{figure}

 	\includegraphics*[width=\linewidth]{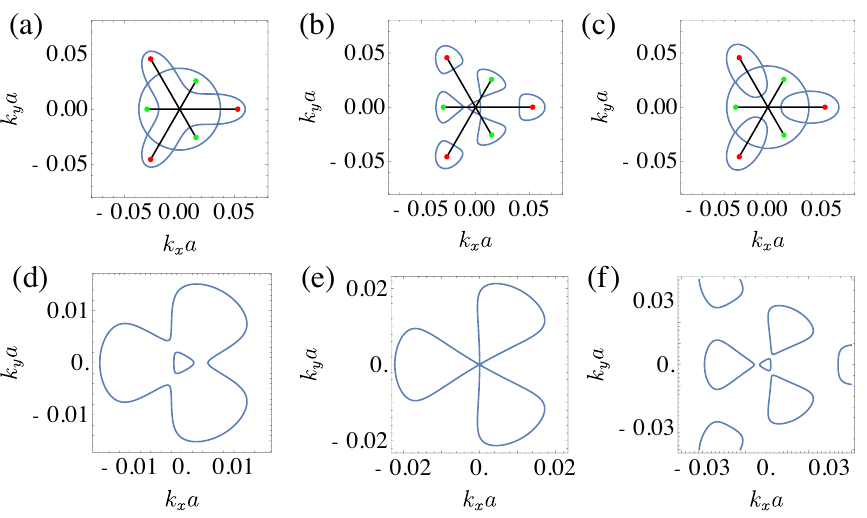}
 	\caption{\label{Fig:LifshitzTransitions} (a)-(c) Fermi contours at the three Lifshitz transitions that happen at fixed $\Delta_1=100$~meV as a function of chemical potential. First Lifshitz transition occurs at the electron side (a), while transitions (b)-(c) happen in the hole band. Positions of the outer and inner gully extrema with distances $Q_1 a=0.053$ and $Q_2a=0.030$ are marked by red and green spots respectively; see also Fig.~\ref{Fig:Bandstructure} that shows band structure at the same value of $\Delta_1$. Panels (d)-(f) show Fermi-contours of the inner hole gullies at fixed value of $\mu=-7$~meV and three different values of electric field, $\Delta_1=40, 60, 80$~meV respectively. In panel (e) the three van Hove singularities of a Lifshitz point join at the origin and form the 'monkey saddle'. 
 	}
 	
 \end{figure}

The formation of gullies necessitates discontinuous change of Fermi-surface topology leading to Lifshitz transitions~\cite{Lifshitz1960} that can be tuned by changing value of the chemical potential $\mu$ at fixed $\Delta_1$. Two such transitions occur on the hole side and one on the electron side. They arise due to the merging of three Fermi pockets from a particular gully into a single Fermi surface as $\mu$ changes. Fermi contours near the transition are shown in Fig.~\ref{Fig:LifshitzTransitions}(a)-(c). The density of states has a van Hove singularity and diverge logarithmically as $\nu(\mu)\sim \log |\mu-\mu_0|$ where $\mu_0$ is the value of chemical potential where Fermi surface contours merge. Observation of these transitions was reported in Ref.~\onlinecite{Rao2018}.
 \begin{figure*}[t]
    \includegraphics*[width=\linewidth]{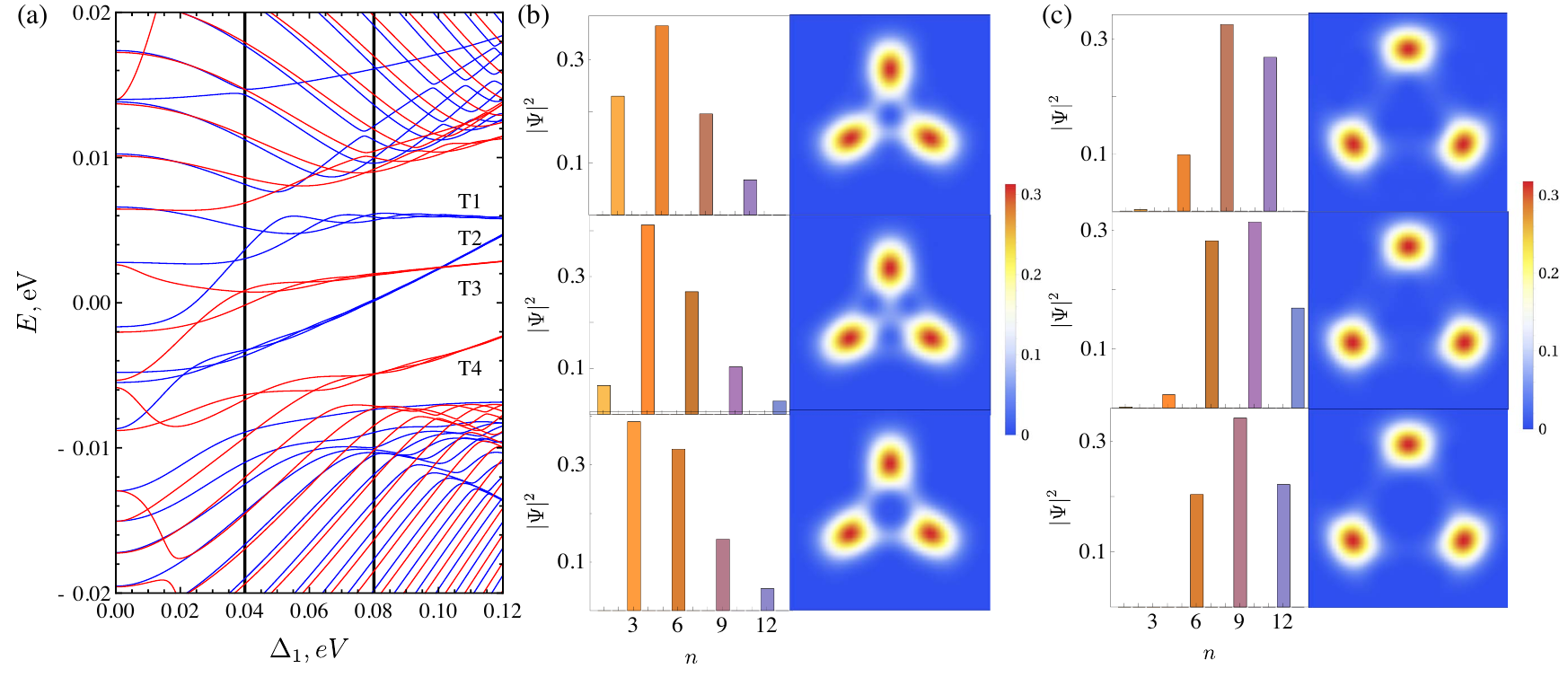}
	\caption{\label{Fig:LLplots} In (a) the spin-degenerate spectrum at $B=1.25$~T is plotted as a function of $\Delta_1$ (left) with blue (red) for LLs in the $K^+$ ($K^-$) valley. The almost equispaced LLs away from neutrality point can be understood semiclassically. This behavior breaks down near the Lifshitz points where LLs intertwine onto each other, forming a set of multiple avoided level crossings. Note that at negative energies there are two such sets, since there are two Lifshitz transitions  in the hole-band. Almost triply-degenerate LLs are formed at large $\Delta_1$. These are labeled T1-4 in order of decreasing energy at $\Delta_1=120$~meV. Intersection of two triplets at $\Delta_1\sim100$~meV correspond to the joining of gullies discussed in Sec.~\ref{Sec:BandStructure}. The dominant wave-function components of three T2 states are concentrated in $B_2$ sublattice and are plotted at (b)  $\Delta_1=40$ and (c) $80$~meV, the LLs from the T2 triplet in each row are given in the order of increasing energy from up to down. The wave function components of T2 triplet shift to higher LL indices as $\Delta_1$ increases. The corresponding density plot visualizes the wave function in real space and shows that the gully distance from origin increases with $\Delta_1$. The axes are scaled by the lattice constant. In (b)-(c), the $x,y$-axis range is $(-600,600)$ in units of lattice constant $a$ with the origin at the $K^+$ point.}
\end{figure*}
 
In addition to Lifshitz transitions, the band structure of ABA graphene has a stronger singularity in the density of states when three van Hove singularities merge at the origin in momentum space, resulting in so-called ``monkey saddle''. Near this point the density of states diverges as a power law. This can be seen by applying the same argument as in the case of bilayer graphene.~\cite{Shtyk2017} We use notations where the saddle point occurs at $k=0$ and at zero energy, $\varepsilon(0)=0$. At this point, the Fermi contours consists of six lines intersecting at the origin, dividing the momentum plane into corresponding regions with alternating signs in energy, see Fig.~\ref{Fig:LifshitzTransitions}(e). From here we deduce that near the origin, $\varepsilon(\bm{k})$ must be proportional to $\cos{3\phi}$ where $\phi$ is the polar angle in the momentum plane. Given that the spectrum itself is not singular, the lowest order terms in the expansion of energy in $k$ must be cubic. In polar coordinates, the expansion reads:
\begin{equation}
\varepsilon(\bm{k}) = \alpha k^3 [\cos{3(\phi-\phi_0)} + C],
\end{equation}
where one can show that constant $C$ satisfies $-1<C<1$. Plugging this expansion into the expression for density of states per unit area,
\begin{equation}\label{DoS}
\nu(\mu) = \frac{g}{2\pi \hbar} \oint_{\varepsilon_{\bm k}=\mu} \frac{dl_{\bm{k}}}{  |\partial \varepsilon/\partial \bm{k}|},
\end{equation} 
(where $g$ is the spin degeneracy),  we obtain that $\nu(\mu)$ diverges as a power law,  
\[
\nu(\mu) \sim |\mu-\mu_0|^{-1/3}, 
\]
where $\mu_0$ is the energy where such monkey saddle occurs. Since such singularity requires the simultaneous meeting of three van Hove singularities, it occurs only at a particular value of electric field $\Delta_{1\text{cr}}\approx 60$~meV. The singularity is located on the hole side spectrum and is shown in Fig.~\ref{Fig:LifshitzTransitions}(d)-(f) where  for comparison, we also show the Fermi surfaces at $\Delta_1$ smaller and larger than $\Delta_{1\text{cr}}$. While this singularity occurs at the energy $\mu_0 \approx -7$~meV and is located within the experimentally accessible range of electric fields, it  seems to be not resolved in the recent experiment reported in Ref.~\onlinecite{Rao2018}.

\section{Landau quantization\label{Sec:MagneticField}}

We now turn to studies of Landau level (LL) spectrum of ABA-stacking graphene. These sets of  LLs were shown in Refs.~\onlinecite{Maksym2013,morimoto}, although previous work did not consider their properties in details. In this section, we focus on the behavior of those gully LLs as a function of $\Delta_1$ at weak magnetic fields. We provide a detailed study of their energy splittings attributed to the magnetic breakdown and visualize the structure of their wave functions. These results are used in Sec.~\ref{Sec:InteractionEffects} to qualitatively understand the role of interaction effects.

To obtain the spectrum, exact diagonalization is performed using the Hamiltonian~(\ref{Eq:Ham}). With a perpendicular external magnetic field $B$, the quasi-momentum operator $\pi$ in Eq.~(\ref{Hamiltonian1}) is replaced by canonical momentum $\Pi = \pi - e (A_x+iA_y)$ where $\bm{A}$ is the vector potential and $e$ is the elementary charge. In the Landau gauge which we adopt throughout this paper, $\Pi$ is the creation (annihilation) operator acting in space of LL indices, $n$ in $K^+$ ($K^-$) valley. Below we present results of numerical study of LL spectrum for $B=1.25$~T and $B=6$~T at different values of transverse electric field, $\Delta_1$. We emphasize that spin degree of freedom is not considered in this section. Indeed, presence of spin simply leads to an approximate additional two-fold degeneracy of all LLs due to small values of Zeeman splitting.

\subsection{Regime of weak magnetic fields \label{Sec:weakfield}}

First, we investigate the LL spectrum at relatively small value of magnetic field, $B=1.25$~T, presented in Fig.~\ref{Fig:LLplots}(a). Most LL features can be understood from the changes of band structure, corresponding to the quasiclassical approximation. Let us review basic changes that were discussed in the literature,\cite{Maksym2013,morimoto} although for different values of magnetic field. We see immediately that the two LLs with energies $E\approx \pm 14$meV at $\Delta_1=0$ that move away from neutrality point with increasing $\Delta_1$ correspond to the tips of the monolayer bands that float away. The approximately equidistant LLs correspond to the remaining two low-energy bands. Their energies decrease as $\Delta_1$ increases since the zero-field low-energy bands move towards the neutrality point with increasing electric field. Lifshitz transition positions are marked by regions where LLs  display numerous anti-crossings that are induced by the tunnelings between different pockets of Fermi surface (magnetic breakdown).

In what follows we focus on the few LLs in vicinity of zero energy which were not studied before. These LLs form groups of three as $\Delta_1$ increases, see Fig.~\ref{Fig:LLplots}(a). Different groups correspond to sets of three gullies related by the $C_3$ symmetry. The four emergent triples of Landau levels are labeled also as T1-4 in correspondence to the labels of gullies in Fig.~\ref{Fig:Bandstructure}. We note, that even at a weak magnetic field, $B=1.25$~T and experimentally accessible values of $\Delta_1$, each gully hosts only three approximately degenerate LLs. Below we concentrate on exploring the structure of the wave function of these triply degenerate LLs. These results will be used in Sec.~\ref{Sec:InteractionEffects} to understand the splitting of their approximate degeneracy by interaction effects. 

The triplet LL states can be described using two natural choices of basis. Analytically, when  gullies are well-separated in the momentum space,  we use a particular set of basis functions centered around each gully, and the inter-gully tunneling is treated as a perturbation. In such ``local basis'' the wave functions in the Landau gauge can be written as:
\begin{equation}\label{Eqn:LL-gully}
\phi_{inX} (x,y) = A_n e^{i \bm{Q}_i\cdot\bm{r}+ i X y/l_B^2-{\gamma^*_i x^2}/{(2l_B^2)}} H_n \bigg(\frac{x}{|\alpha_i| l_B}\bigg) \chi_{in},
\end{equation}
where $n, i$ correspond to the LL index and gully index respectively, $X$ is the guiding center coordinate and $\chi_{in}$ is the fixed pseudospinor in layers and sublattices. $l_B = \surd(\hbar c/eB)$ is the magnetic length, $\bm{Q}_i$ is the distance from the origin to the center of the given gully in momentum space and $H_n(x)$ is the $n$-th Hermite polynomial. Constants $\alpha_i$ and $\gamma_i$ characterize  anisotropy of the gully, and $A_n$ is the normalization factor; their definitions and derivation of Eq.~(\ref{Eqn:LL-gully}) are delegated to Appendix~\ref{Appendix:FormFactors}. In the local basis, each gully contains only the $n=0$ LL, since gullies are fairly shallow in the physical range of $\Delta_1$ and higher LLs would only appear at smaller $B$. Hence, in what follows we discuss only wave functions with $n=0$ in the local basis.

\begin{figure*}[t]
	\includegraphics*[width=\linewidth]{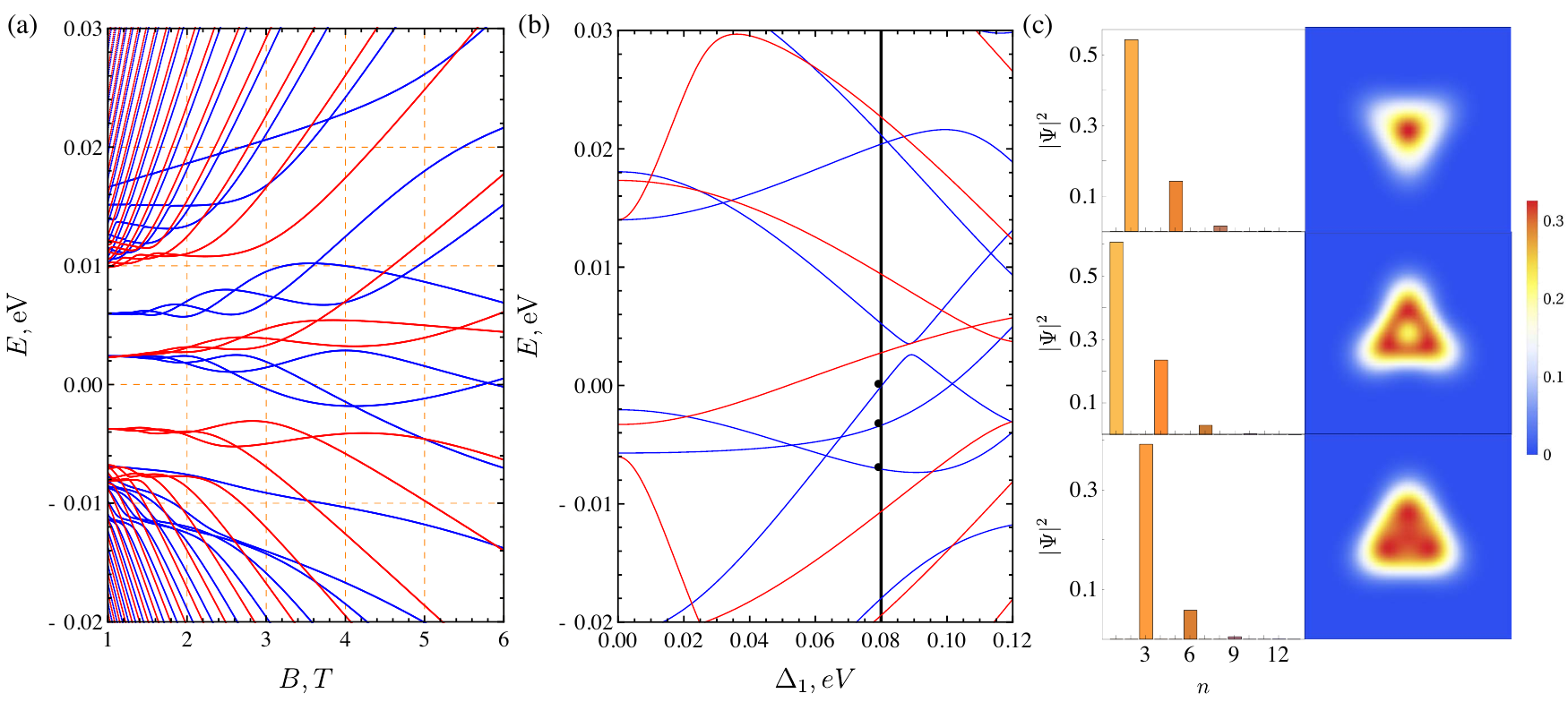}
	\caption{\label{Fig:StrongField} (a) LL spectrum at $\Delta_1=100$~meV plotted as a function of $B$ shows that triplet degeneracies are rapidly lifted with increasing magnetic field.  (b) LL spectrum at $B=6$~T is plotted as a function of $\Delta_1$ illustrates that triplets lose their gully character as is manifested by avoided crossings between different Landau levels. (c) Wave-function components on the $B_2$ sublattice  of LLs that formally belong to T2 triplet [marked by dots in panel (b)] show that eigenstates are concentrated near lower LL indices compared to Fig.~\ref{Fig:LLplots}(c) due to reduced magnetic length. Magnetic field is $B=6$~T, electric field is $\Delta_1=80$~meV. The $x,y$-axis range is $(-600,600)$ in units of lattice constants with the origin at the $K^+$ point.
}
\end{figure*}

However, in our system the gullies are in general close to each other in momentum space, so that inter-gully tunneling cannot be neglected already at very moderate values of magnetic field. Therefore, while the local basis is convenient for analytical considerations, in the limit of significant inter-gully tunneling it is more natural to consider the ``global basis'' which expands triplet LLs in isotropic LL wave functions centered at $K^{\pm}$ on a given layer and sublattice $\alpha$,
\begin{equation} \label{Eq:LL-global}
\psi_{\alpha nX} (x,y)= A_n e^{i X y/l_B^2-x^2/(2l^2_B)}H_n\left(x/l_B\right)\chi_\alpha,
\end{equation}
where $\chi_\alpha$ is the pseudospinor corresponding to layer and sublattice $\alpha$ that has six possible values, $A_1,B_1,A_2,B_2,A_3,B_3$. This is the basis that is used by numerical diagonalization, and in Fig.~\ref{Fig:LLplots}(b)-(c) we illustrate the structure of wave functions of T2 LL in this basis around $K^+$ point. Since LLs in triplet T2 are concentrated on $B_2$ sublattice, the wave function of T2 can be approximated as 
\be\label{Eq:T2WF}
\Psi_{\text{T2},X}(x,y)  \approx \sum_n c_{{B_2},n}\psi_{{B_2}nX}(x,y),
\ee
and we show in the bar chart $|c_{{B_2},n}|^2$ only. LL from other triplets are concentrated on $B_2$ for triplet T1 and on $A_1, B_1, A_3, B_3$ sublattices for triplets T3 and T4.  Moreover we note that wave function coefficients $c_{\alpha n}$ are non-zero for LL indices $n$ that differ by multiples of three. This feature is a consequence of the $C_3$ invariance of the Hamiltonian. This symmetry enforces the wave function to be a coherent superposition of different gullies.

From numerical results we observe that wave function has  $|c_{B_2,n}|^2$ peaked at some $n_{\rm max}$ that is generally not close to zero. This aspect of the LL wave function in the global basis can be understood using the simple analytic structure of the wave function in the local gully basis. Indeed, the coefficients $c_{B_2,n}$ can be calculated as overlaps between basis wave functions $\phi_{i0X}(x,y)$ and $\psi_{B_2 n X}(x,y)$ from Eqs.~(\ref{Eqn:LL-gully})-(\ref{Eq:LL-global}). The basis functions in Eq.~(\ref{Eqn:LL-gully}) have displaced origin, and presence of the ``boost operator'' $\exp(i\bm{Q}_i\cdot\bm{r})$ causes expansion coefficients to be peaked at $n_{\rm max}\sim \sqrt{Ql_B}$. Thus as centers of the gullies move further away from $K^\pm$ point, and  $Q$ increases,  triplet level components are concentrated at higher LL indices, cf.~panels (b)-(c) of Fig.~\ref{Fig:LLplots}. 

The same trend is also apparent in the plots of the real space probability density using quasi-classical wave-functions. The quasiclassical wave functions are obtained from an convolution of a basis states $\psi_{\alpha n X}(x,y)$~(indices $\alpha, n $ are fixed) with a Gaussian envelope function $C_X$ that maximally localizes the resulting wave packet, see Appendix~\ref{Appendix:QuasiClassicalWF}. The probability density can be understood intuitively as concentrating around the classical cyclotron orbit. Since, in a magnetic field, real space quasi-classical trajectories of electrons are obtained from constant energy contours in moment space by $\pi/2$ rotations and rescaling by $l_B^2$, probability densities reflect gully positions. Comparing density plots in panels (b)-(c) in Fig.~\ref{Fig:LLplots} we observe that the dominant weight is displaced further away from $K$ point with increasing $\Delta_1$.

Finally, we return to the discussion of the splitting of three-fold degeneracy of the triplets by magnetic breakdown. The process of tunneling between gullies is automatically taken into account by exact diagonalization, hence the individual LLs in triplets T1-T4 in Fig.~\ref{Fig:LLplots}(a) oscillate with respect to each other. On the other hand, at the level of analytical (gully) LL wave functions, the effect can be taken into account by introducing a tunneling between triplets that has a form
\begin{equation}\label{Eqn:Tunelling}
H_T=\begin{pmatrix}
0 & t & t^* \\ t^* & 0 & t \\ t & t^*&0
\end{pmatrix}
\end{equation}
in the local basis of triplet states. Such tunneling breaks the triplet degeneracy $\epsilon_1 = \bar \epsilon+2 |t| \cos \phi$, $\epsilon_{2,3} =\bar \epsilon -2 |t|\cos (\phi \pm 2\pi/3)$ where $\phi$ is the phase of $t$ and $\bar \epsilon$ is the LLs energy without tunneling. The effective tunneling can be calculated using analytic framework of Ref.~\onlinecite{Alexandradinata2018}. Its magnitude can be estimated as~\cite{Alexandradinata2018} 
\begin{equation}\label{Eq:t-est}
|t|\sim \omega_0 \exp [-\frac{\pi}{8} Q^2 l_B^2 \sqrt{\frac{m_y}{m_x}}].
\end{equation} 
$m_{x,y}$ is the effective mass with principle $x$-axis joining two gullies, $Q$ is the magnitude of the classical forbidden momentum range. $\omega_0$ is the cyclotron frequency associated with the motion on the semiclassical orbit. In the limit of large $\Delta_1$ and gully separation, Eq.~(\ref{Eq:t-est}) becomes
\begin{equation}\label{Eq:t-est1}
|t|\sim \omega_0 \exp(-C \Delta^2_1/B).
\end{equation} 
$C$ is a constant that depends on band geometry and tight-binding parameters. We expect that $\omega_0$ varies slowly with $\Delta_1$, thus in the limit of weak magnetic breakdown, the splitting between triplets is expected to be exponentially sensitive to $\Delta_1$.

\subsection{Regime of strong magnetic fields}

In this section we follow the fate of the low energy triply degenerate LLs as the magnetic field strength is increased. Figure~\ref{Fig:StrongField}(a) shows the spectrum as a function of $B$ at $\Delta_1=100$~meV. Since the band structure is determined by $\Delta_1$, at small values of $B$ only the triplet cyclotron gaps change. Upon increasing magnetic field, amplitude of splitting of triplet LL energy increases due to increased tunneling. At sufficiently large $B$, the inter-gully tunneling becomes so strong that `triplet' states entirely lose their gully character due to magnetic breakdown between different gullies. 

This can be visualized by plotting the energy spectrum at $B=6$\,T as a function of $\Delta_1$, see Fig.~\ref{Fig:StrongField}(b) where the triplet energy splittings become larger than cyclotron gaps between different triplets. Magnetic breakdown effects become strong  when $\delta k  l_B\sim1$, where $\delta k$ is the smallest distance between two Fermi contours corresponding to the semiclassical gully LLs. At large $\Delta_1$ or small $B$, when the size of Fermi surface of a given LL is small comparing to inter-gully distances $\Delta Q$, $\delta k \approx \Delta Q\sim \Delta_1$, and the magnetic field corresponding to the onset of magnetic breakdown increases quadratically with electric field, $B\sim \Delta_1^2$, see Eq.~(\ref{Eq:t-est1}).

Finally, we illustrate the structure of LL wave functions in the regime of strong magnetic breakdown  in Fig.~\ref{Fig:StrongField}(c). We concentrate on the structure of wave function components and probability densities of T2 at $B=6$\,T. From the plots of the real-space probability density we conclude that LLs are concentrated near the origin and look qualitatively different from the regime of small $B$;  see Fig.~\ref{Fig:LLplots}(c). However, the ``mod 3'' pattern in expansion coefficients described in the Sec.~\ref{Sec:weakfield} still persists. This feature can be potentially used for the tunability of interactions in the regime of strong electric fields, helping to realize interesting fractional quantum Hall states and phase transitions via tunability of form-factors.~\cite{Papic2012}

\section{Interaction Effects\label{Sec:InteractionEffects}}
As we discussed in previous section, in the absence of interactions and magnetic breakdown, the single-particle degenerate eigenstates are linear superpositions of gully states of the same LL index that realize three irreducible $C_3$ representations. Electron-electron interactions are expected to alter this picture considerably, potentially resulting in symmetry-breaking ground states. In this section we focus on interaction effects in the case of $\nu=1$ filling. After a brief review of variational Hartree-Fock approximation that uses the local gully basis, we use them to present analytical approximation at large inter-gully distances. Then we perform numerical study of Hartree-Fock states that fully incorporates effect of intergully exchanges that are partially neglected in the analytic treatment. In both cases we find the phase transition between gully polarized state and gully coherent state to be of the first order. 

In this section, the magnetic length $l_B$ is set to one. Moreover, as we discuss below, we largely ignore spin degree of freedom. Exchange interactions favor complete spin polarization whereas Zeeman field fixes the direction of this polarization.

\subsection{Hartree-Fock approximation in gully basis}

To set up analytical calculations, we first discuss the HF approximation in the basis of gully LLs. We only consider one set of $C_3$ symmetric gullies. Thus the total Hamiltonian has the form:
\begin{equation}\label{Eq:hfull}
H = H_0+U+H_{\text{ZM}},
\end{equation}
where $H_0$ is the spin-degenerate single particle Hamiltonian and $H_{\text{ZM}}$ is the Zeeman term. The interaction term is given by:
\be\label{Eqn:Interactions}
U = \frac{1}{2} \int d^2\bm{r}_1d^2\bm{r}_2 U(\bm{r}_1-\bm{r}_2)\Psi^\dagger(\bm{r}_1) \Psi^\dagger(\bm{r}_2) \Psi(\bm{r}_2) \Psi(\bm{r}_1) ,
\ee
where $U(\mathbf{r})$ is the two-dimensional Coulomb potential. 

In what follows we are interested in the $\nu=1$ ground state. The single particle Hamiltonian, $H_0$, does not depend on spin, whereas exchange terms in $U(\mathbf{r})$ favor electron polarization. Therefore we assume that all spin is aligned with the magnetic field to minimize the Zeeman energy, $H_{\text{ZM}}$. This allows us to omit spin degrees of freedom in what follows.~\cite{Comment1}

In Sec.~\ref{SubSec:Analytics}, we consider large inter-gully distances and weak magnetic field, such that the energy splitting between different LLs from the same triplet is negligible. Then both $H_0$ and $H_{\text{ZM}}$ in Eq.~(\ref{Eq:hfull}) give an overall energy shift and can be ignored. In order to treat the remaining interaction term, and find the ground state at $\nu =1 $ we use the HF approximation which finds the best wave function in the variational manifold. We write the $\Psi$ operators in Eq.~(\ref{Eqn:Interactions}) in the second quantized language,
\be\label{Eqn:PsiOperator}
\Psi(\bm{r}) = \sum_{i,n,X} \phi_{inX} (\bm{r}) a_{inX},
\ee
where $a_{in X}$ is the electron annihilation operator and the basis wave function $\phi_{i nX}(\bm{r})$ is given in Eq.~(\ref{Eqn:LL-gully}). The HF variational wave function for a given LL $n$,
\be\label{Eq:variational-wf}
 \ket{n,\{c_i\}} = \prod_X \bigg(\sum_{i=1}^3c_i a_{inX}^\dagger \bigg)\ket{0},
\ee
depends on three complex parameters, $c_i$, that specify amplitudes of degenerate gully states. 

Using this variational wave function we calculate the expectation value of interaction term and optimize it over values of $c_i$. In the process of calculation we use expectation values of creation and annihilation operators. For instance, two-operator expectation value reads,
\[
\langle n,\{c_i\}| a^\dagger_{i_1n_1X_1}a^{}_{i_2n_2X_2}|n,\{c_i\}\rangle = c^*_{i_1}c^{}_{i_2} \delta_{n_1n_2}\delta_{X_1X_2}.
\]
Assuming that the density-density term is neutralized by a positive charge background, we obtain the exchange energy as a quartic polynomial in $c_i$:
\begin{eqnarray}\label{ExchangeEnergy}
U_{\text{ex}}&=&-\frac{1}{2} \sum_{i_1,i_2,i_3,i_4=1}^3 J^{(n)}_{i_1i_4,i_2i_3} c^*_{i_1}c^*_{i_2}c^{}_{i_3}c^{}_{i_4} , \\
J^{(n)}_{i_1i_4,i_2i_3} &=& \int U(q) F^{nn}_{i_1 i_4}(-q) F^{nn}_{i_2i_3}(q) \frac{d^2q}{(2\pi)^2}, \label{Eq:exch}
\end{eqnarray}
where $F^{nm}_{i_1i_2}(q)$ are form factors derived in Appendix~\ref{Appendix:FormFactors}.  $U(q) = 2 \pi e^2/[q \varepsilon(q)]$ is the Fourier transform of the Coulomb potential where $\varepsilon(q)$ is the dielectric function describing screening. For $B=1.25$~T and within the accessible range of $\Delta_1$, T1-T4 correspond to the zeroth LL in each gully, therefore in Eq.~(\ref{ExchangeEnergy}) only such wave-functions are considered and we omit LL index $n=0$ in the following. 

The exchange integrals $J_{i_1i_4,i_2i_3}$ in Eq.~(\ref{Eq:exch}) characterize interactions between LLs and can be constrained using lattice symmetries. Note, that one can neglect the dependence of $U(q)$ on the structure of wave functions in layer space. Indeed, the interactions between layers introduce an additional factor $\exp(-qd)$ where $d$ is the layer distance [see Eq.~(\ref{Potential}) in Appendix], and  the important range of integration in Eq.~(\ref{Eq:exch}) is $ql_B\sim 1$. In graphene, adjacent layer distance is $d=0.335~\text{nm}$ and $l_B\gg d$ always holds, therefore layer structure of the wave function can be neglected in Eq.~(\ref{ExchangeEnergy}).

The expectation value of interaction energy $U$ is minimized with respect to $c_i$ to find the ground state. Minimization of Eq.~(\ref{ExchangeEnergy}) is in general not possible analytically. However, as will be shown in the next subsection, the situation is considerably simplified in the limit of large gully distance and small anisotropy. This allows us to derive analytical results that illustrate qualitatively the evolution of ground state as a function of gully distance which is tuned by $\Delta_1$.

\subsection{Analytical results in gully basis}\label{SubSec:Analytics}

From Eq.~(\ref{ExchangeEnergy}), we see that each pair of indices in $J_{i_1i_4,i_2i_3}$ refers to matrix elements taken between the two gully states. Transitions between different gully states are suppressed exponentially by the momentum space distance $Q$ between centers of the two gullies in the reciprocal space, see Appendix~\ref{Appendix:FormFactors}. In the limit of large $Q$, exchange integrals of the form $J_{ii,kk}$ are dominant and, neglecting  scattering between different gullies, Eq.~(\ref{ExchangeEnergy}) simplifies into:
\begin{equation}\label{ExchangeEnergy1}
U_{\text{ex}}=-\frac{1}{2} \sum_{i_1,i_2} |c_{i_1}|^2|c_{i_2}|^2 J_{i_1i_1,i_2i_2}.
\end{equation}
This is also the limit considered in, i.e. Refs.~\onlinecite{Sodemann2017,Cheung2018,Abanin2010,Kumar2013,Kumar2016}. Due to $C_3$ symmetry, $J_{ii,kk} = J_{ik}$ have same values for all diagonal elements $J_0$ and all off-diagonal ones $J_1$. Then the ground state is given by minimizing 
\begin{equation}\label{Eq:Exch-simple}
U_{\text{ex}} =- \frac{1}{2}\bigg(J_0\sum_i |c_i|^4+J_1\sum_{i\ne k}|c_i|^2|c_k|^2\bigg)
\end{equation}
that follows from Eq.~(\ref{ExchangeEnergy1}). Eq.~(\ref{Eq:Exch-simple}) is minimized by the fully gully-polarized $c_i=1$ state, provided that the gullies are anisotropic, which leads to $J_0> J_1$.~\cite{Abanin2010,Sodemann2017} When $J_1=J_0$, the system has SU$(3)$ symmetry in the space of gully states and $U_{\text{ex}}=-J_0$ for any values of ${c_i}$.
 
However the maximal gully polarization cannot persist when gullies become close to each other. Indeed, in the opposite limit of very small inter-gully distance, we expect all terms in Eq.~(\ref{ExchangeEnergy1})  to be of comparable magnitude $J$. In this limit, the HF ground state becomes a coherent superposition of $N=3$ gully states. This can be seen from the following argument: the coherent state has $c_i \sim 1/\sqrt{N}$ and $U_{\text{ex}} \sim - J \sum_{i_1,i_2,i_3,i_4} (1/\sqrt{N})^4 \sim -JN^2$, which is energetically favorable to gully polarized state with $U_{\text{ex}} \sim -J$. 

\begin{figure}[t]
\includegraphics[width=.8\linewidth]{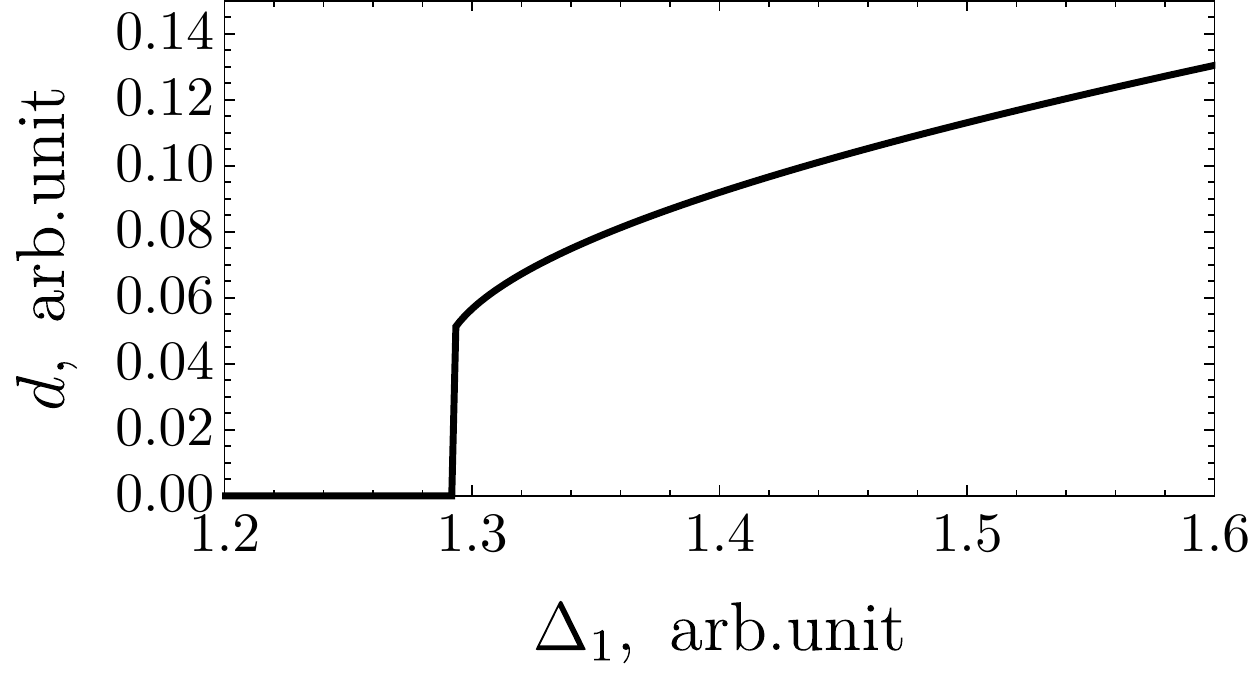}
\caption{\label{Fig:Dipole} The dipole moment calculated from Eq.~(\ref{Eqn:Dipole}) has a jump as a function of $\Delta_1$ . The state is obtained by minimizing the Eq.~(\ref{ExchangeEnergy3}) with couplings set to $J_0-J_1=0.3$ and $J_2 = \exp(-\Delta_1^2)$.}
\end{figure}

By the above argument, the maximal gully-polarization is expected to break down upon increasing inter-gully scattering. To investigate this transition in greater details we include first order corrections due to inter-gully scattering in addition to terms in Eq.~(\ref{Eq:Exch-simple}). These are terms of the form $J_{ii,kl}|c_{i}|^2c^*_{k}c_{l}$ with $k\ne l$. In the limit of small anisotropy, we could regard $J_{ii,kl}$ as calculated with the isotropic gully wave-functions and the only parameter is $Q_{lk}$, the magnitude of momentum transfer between gullies $l$ and $k$. Thus, to a first approximation, all $J_{ii,kl} = J_2$ can be regarded as equal due to rotational symmetry.  From Eq.~(\ref{AnisotropicFormFactors}) we see that $J_2 \sim J_0\exp(-Q^2/4)$, where $Q$ is measured in units of inverse magnetic length. We can also neglect the single particle energy splitting due to tunneling. It is of the magnitude $|t|$ and, from Eq.~(\ref{Eq:t-est}), $|t|\sim \omega_0\exp(-\pi Q^2/8)\sim \omega_0 (J_2/J_0)^{1.57}\ll J_2$. Thus the resulting exchange energy reads:
\begin{multline}\label{ExchangeEnergy3}
U_{\text{ex}} = - \frac{1}{2} \bigg( J_0\sum_i |c_i|^4 +J_1 \sum_{i\ne k}|c_i|^2 |c_k|^2\bigg)\\ -J_2\sum_{i\ne m,k}c^*_i c_m |c_k|^2.
\end{multline}

The nature of the ground state that minimizes Eq.~(\ref{ExchangeEnergy3}) depends on the value of dimensionless parameter $\kappa = J_2/(J_0-J_1)$. For small $\kappa$, the ground state is still strongly gully polarized but with non-zero components in all gully basis. At a certain critical $\kappa_c\approx 0.25$, the ground state becomes fully gully coherent with $c_1=c_2=c_3=1/\sqrt{3}$ and the phase transition is of the first order (see detailed discussion in Appendix~\ref{Appendix:Proof}).

To characterize the nature of $C_3$ symmetry breaking transitions, the most natural order parameter is the dipole moment. To understand qualitatively its behavior, we use the quasi-classical dipole moment which is the time-averaged position vector of classical cyclotron motion.~\cite{Sodemann2017} Here we approximate it as the sum of real space position vectors $\bm{r}_i$ of each gully multiplied by their weight in the wave function, $|c_i|^2$ and LL degeneracy $N_\phi = eB/2\pi \hbar$:
\begin{equation}\label{Eqn:Dipole}
\bm{d} =e N_\phi \sum_i |c_i|^2 \bm{r}_i.
\end{equation}

We now consider the dipole moment of our HF state using Eq.~(\ref{Eqn:Dipole}). Since, in a magnetic field, real space quasi-classical trajectories of electrons are rotated moment space orbit, the magnitude of ${r}_i$ is proportional to ${Q}_i\propto \Delta_1$. The behavior of the dipole moment across the phase transition is shown schematically in Fig.~\ref{Fig:Dipole}, where the dipole moment $d$ is plotted as a function of $\Delta_1$ which controls the suppression of inter-gully scattering. The discontinuous jump reveals the first order phase transition where spontaneous gully polarization develops.

\subsection{Numerical Results for TLG triplets}\label{SubSec:Numerics}

In the analytical treatment presented above, we ignored gully anisotropy and single particle energy splitting due to magnetic tunneling. However, in realistic systems, the anisotropy of gullies cannot be regarded as a small perturbation. Also, magnetic breakdown is already significant even at the very weak fields. Hence, below we investigate numerically the nature of the $\nu=1$ HF ground state, using LL coefficients obtained from exact diagonalization outlined in Sec.~\ref{Sec:MagneticField}. Since we use the exact Hamiltonian expanded near the $K^{\pm}$ points, this procedure automatically takes into account all the tunneling and anisotropic effects. While these perturbations may change the location of phase transition where spontaneous gully polarization develops, we observe that it remains to be of the first order. For numerical HF calculations, we follow the approach outlined in Ref.~\onlinecite{Zhang2012} and use an interpolation formula for the dielectric function $\varepsilon(q)$ to take account of screening.~\cite{Papic2014} Details of our numerical simulation and choice of screening are discussed in Appendix \ref{Appendix:HartreeFock}. In addition, we discuss the qualitative effect of screening in the end of this section. 

\begin{figure}[t]
\includegraphics[width=0.99\linewidth]{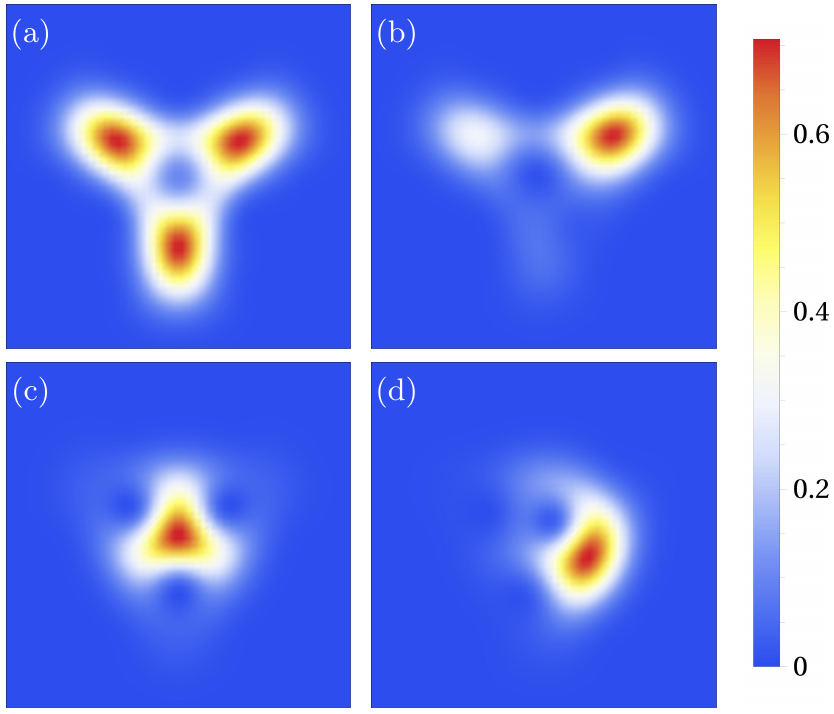}
	\caption{\label{Fig:HFProbability} Real space probability densities for T3 filling factor 1 HF ground states are plotted at (a) $\Delta_1=40$~meV; (b) $\Delta_1=50$~meV. HF ground state in (a) coincides with the single-particle eigenstate while in (b) it exhibits gully polarization.  Real space probability densities for T4 filling factor 1 HF ground states are plotted at (c) $\Delta_1=70$~meV; (d) $\Delta_1=80$~meV. In (c) the HF state coincides with the single-particle one . In (d) it is strongly polarized into one of the gullies. The $x,y$-axis range is $(-600,600)$ in units of lattice constant $a$ with the origin at the $K^-$ point.}
\end{figure}

We apply the HF procedure to triplets T1-T4 (see Fig.~\ref{Fig:LLplots}) in the range of values of $\Delta_1$. We note that setup when $\Delta_1$ is a tuning parameter is more natural, since changing magnetic field would lead to a varying filling factor. Before discussing generic results, we illustrate the wave functions deep in the gully polarized and symmetric phases; all electron spins are up. For instance, the HF calculation for T$3$ at $\Delta_1=50$~meV reveals gully polarized state, whereas at $\Delta_1=40$~meV, the HF groundstate coincides with the single particle state; see Fig.~\ref{Fig:HFProbability}(a)-(b) for the wave function visualization. Another example is provided by HF calculations on T4 that has larger anisotropy as can be seen from Fig.~\ref{Fig:Gullies}(b). As shown in Fig.~\ref{Fig:HFProbability}(c)-(d), at $\Delta_1=70$~meV, the HF eigenstate coincides with the single particle state and at $\Delta_1=80$~meV, the HF state becomes gully-polarized. Symmetry breaking occurs at much closer inter-gully distance, which  is consistent with analytic arguments in Sec. \ref{SubSec:Analytics}. Indeed, the high anisotropy of pockets in T4 reduces the magnitude of inter-gully scattering form factors in exchange integrals.

\begin{figure}[t]

\includegraphics[width=\linewidth]{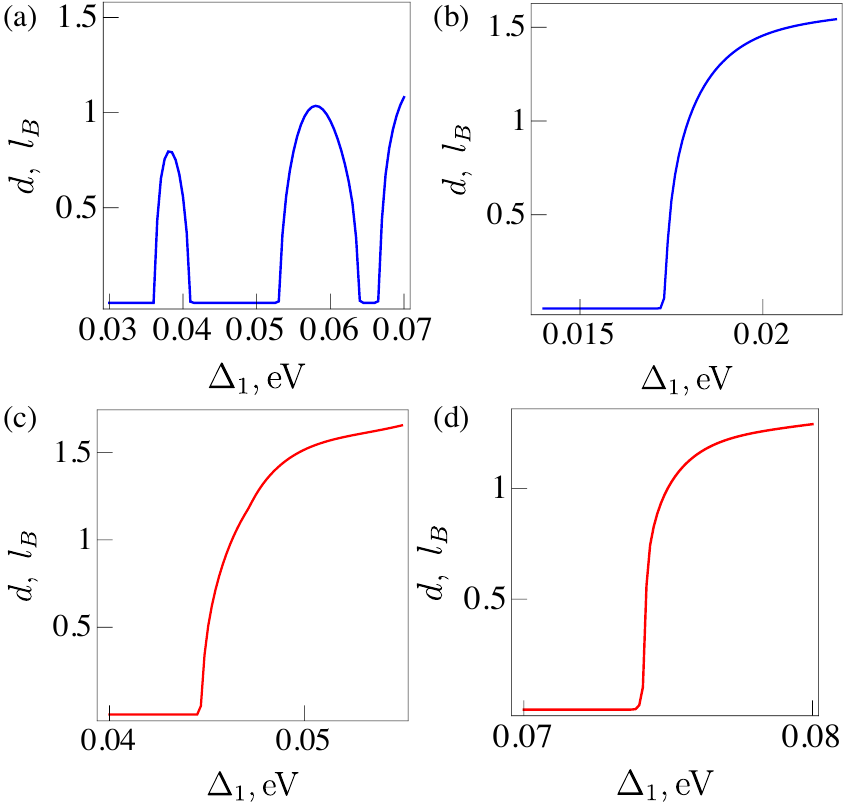}
	\caption{\label{Fig:HFDipole}  Panels (a)-(d) show discontinuous change of dipole moments as a function of $\Delta_1$ for T1-4 respectively. In particular, the oscillating behavior of T1 dipole moment is due to the oscillation of single particle energies of triplets with $\Delta_1$. }
\end{figure}

Gully polarized states can be accessed experimentally by measuring dipole moment. The explicit expression for the expectation value of the dipole moment is given in Appendix~\ref{Appendix:DipoleMoment}. Fig.~\ref{Fig:HFDipole} shows the dipole moment $d$ calculated for triplets T1-4 as a function of $\Delta_1$. It might seem from Fig.~\ref{Fig:HFDipole} that, for all triplets and just below the critical $\Delta_1$ value, $d$ has a small and smooth initial increase, then undergoes a discontinuous change at the critical point. We believe this is due to the slowness of numerical convergence near the critical point: while our iteration number is sufficient for convergence of HF loops for most $\Delta_1$ values, near the critical point, the numerics fail to converge and the initial small $d$ values decrease considerably with more iterations. Thus, the transition is still expected to be first order. In Appendix \ref{Appendix:DipoleMoment}, we present additional arguments in support of this statement. 

We see from Fig.~\ref{Fig:HFDipole} that, while T2-4 exhibit one single discontinuity in dipole moment, several transitions occur for T1. We attribute the alternating appearance and vanishing of dipole moment in T1 the large splitting of energies of single particle LLs in this triplet due to magnetic breakdown~(see Fig.~\ref{Fig:LLplots}). In addition, the development of gully polarization for T2 happens at lower value of $\Delta_{1}$. This can be explained by the weaker tunneling between different pockets in T2 and, consequently, smaller splitting of degeneracy. We also conclude from comparing values of $\Delta_{1}$ where transition occurs and Fig.~\ref{Fig:Gullies}(a), that gully polarization sets in for T2 at greater inter-gully distance compared to T1-4. This is consistent with the smaller anisotropy of triplets T2-3 compared to T1-4, as is shown in Fig.~\ref{Fig:Gullies}(b).

Finally, we comment on the form of dielectric function $\varepsilon(q)$ used in numerical HF calculations above. We have used an interpolation formula for $\varepsilon(q)$ in the limit of strong screening which is suggested by measurements in Ref.~\onlinecite{Rao2018}. This approach provides an order of magnitude estimate, that qualitatively agrees with the range of $\Delta_1$ where experiment begins to resolve the large gap between triplets.~\cite{Rao2018} The realistic fully microscopic  calculation of screening is challenging as it requires the knowledge of microscopic interactions in the system and incorporation of effects of filled Landau levels into screening. However, we can understand the overall effect of weaker screening (provided that the overall scale of interactions remains fixed) qualitatively. Generally we expect the Coulomb potential to remain unscreened at short distances (large momenta) and suppressed at larger distances (small momenta). Since inter-gully scattering destroys gully-polarization, qualitatively we expect the strong screening to favor gully-coherent states. Indeed, stronger screening reduces the relative ratio between exchange integrals for intra-gully scattering and inter-gully exchange and scattering. In the opposite limit of weak screening, we expect gully polarization to set in at even smaller~$\Delta_1$.

\section{Summary and Outlook\label{Sec:Discussion}}

In this work we focused on the role of interaction effects in the ABA-stacking trilayer graphene in presence of strong transverse electric (displacement) field and magnetic field. In this regime the single-particle band structure is characterized by new emergent Dirac points --- gullies --- that were theoretically predicted in Refs.~\onlinecite{Maksym2013,morimoto} and recently observed experimentally in Ref.~\onlinecite{Rao2018}. First, we characterized properties of gullies in a non-interacting band structure using the recently obtained set of tight-binding parameters. In addition, we identify multiple Lifshitz transitions and higher-order singularity of ``monkey-saddle'' type that can be tuned by chemical potential or displacement field.

In presence of weak transverse magnetic field these gullies lead to a three-fold degenerate Landau levels observed earlier.~\cite{Maksym2013,morimoto} Moving beyond the results obtained earlier,  we considered the structure of wave functions of these three-fold degenerate LLs. The understanding of the structure of wave functions and effects of magnetic breakdown was used to understand the lifting of the three-fold degeneracy by interaction effects. We considered the effect of interactions on three-fold degenerate sets of LLs at integer filling $\nu=1$ within the Hartree-Fock approximation. In a case of strong displacement field corresponding to gullies being well-separated in reciprocal space, interactions favor states with gully polarization that break $C_3$ rotational symmetry. However, at stronger magnetic fields or smaller values of displacement field we find a gully coherent state. Within Hartree-Fock approximation, the breakdown of gully coherence happens via first-order phase transition that is characterized by an emergence of non-zero expectation value of the dipole moment.

Our results suggest that multi-layer graphene is a promising platform for investigating interaction effects. Without magnetic field, singularities in density of states at Lifshitz points and monkey saddle may potentially host novel interaction-induced states. The particularly promising region to search for such states is between two Lifshitz points on the hole side, where the experimental quantum capacitance measurements confirmed the existence of the region with particularly high density of states.~\cite{Rao2018} 

In presence of magnetic field, the ABA graphene is expected to host interaction-driven symmetry broken states similarly to the case of other multi-valley platforms such as SnTe-(111),~\cite{Li2016} PbTe-(111),~\cite{Chitta2006} and Bi-(111).~\cite{Koroteev2004} However, the ABA trilayer graphene enjoys additional tunability compared to other platforms. The displacement field changes the distance between gullies in reciprocal space. As we discussed above, there exists the first order phase transition separating the gully-coherent and partially gully-polarized ground states. This phase transition can be tuned by the strength of displacement field. It is characterized by emergence of non-zero dipole moment in the gully polarized state. Thus, biased ABA trilayer graphene allows for observation of phase transition that may be inaccessible in other multi-valley materials where gullies are well-separated.

Recent realization of extremely high-quality ABA graphene encapsulated in hBN with graphite gates, Ref.~\onlinecite{Rao2018}, provides the first step towards observation of the physics discussed above. Indeed, the experimental data reported in Ref.~\onlinecite{Rao2018} strongly suggest existence of symmetry broken states at integer fillings of gully LLs. However, establishing the nature of these states requires further investigation. On the experimental side, it would be interesting to perform transport measurements on these states that can be potentially capable of detecting anisotropy that originates from $C_3$ symmetry breaking. Alternatively, scanning tunneling microscopy (STM) can be potentially used to directly visualize LL wavefunction profiles pinned by local impurities.~\cite{Papic2018,Sid2019} Therefore STM could potentially be useful for probing symmetry-breaking states in our system, although more detailed study is needed to understand feasibility of such setup.

Theoretically, the dielectric function is an important ingredient used in the Hartree-Fock calculations, that is challenging to calculate realistically. Hence predicting the exact location of the phase transitions theoretically remains challenging. As we discussed in Sec.~\ref{SubSec:Numerics}, qualitative effect of weaker screening (provided that overall scale of interactions stays the same) is the shift of gully polarization transition to smaller values of displacement field. Experimentally, this may enable tuning the location of phase transition via changing the dielectric thickness or even using suspended samples.\cite{Nam2018}

Finally, we discuss the physics beyond the Hartree-Fock approximation considered in this work. The analytical considerations in Sec.~\ref{Sec:InteractionEffects} presented a model with an approximate SU$(3)$ symmetry in the space of gully states, which is explicitly broken by small anisotropy and inter-gully scattering. Provided that symmetry breaking is weak, the disorder may lead to a presence of domains with different order parameters. Moreover, the low energy excitations are given by `gully-wave' Goldstone modes may influence the physical properties of the system. Both of these ingredients are beyond the na\"ive Hartree-Fock approximation with spatially uniform order parameter adopted here. The disorder and Goldstone mode effects were considered for two-valley systems with approximate SU$(2)$ valley-symmetry.~\cite{Rasolt1986,Abanin2010,Kumar2013,Kumar2016} In the SU$(2)$ case, valley configurations can be formally characterized as spin states and mapped to an effective O$(3)$ non-linear sigma-model. The model predicts the existence of charged topological excitations at domain walls that separate different valley coherent configurations.~\cite{Abanin2010,Kumar2013,Kumar2016} In addition, Ref.~\onlinecite{Abanin2010} suggests that weak disorder might be sufficient to destroy macroscopic gully polarization but preserve gapped quantum Hall state. We expect similar topological defects to be present in our system. Thus, the study of effective theory for Goldstone modes in the SU$(3)$ case and understanding of disorder effect remains an interesting open question. 

\section*{Acknowledgements}
We thank A. Young, S. Zibrov, and C. Kometter for the experimental collaboration that attracted our attention to this problem. We acknowledge useful discussions with A. Goremykina, A. Michailidis, Z. Papic, and, especially, S. Parameswaran. 

\appendix
\section{Band structure without displacement field}\label{Appendix:Bandstructure}
We use the Slonczewski-Weiss-McClure parametrization of the tight-binding model introduced in Ref.~\onlinecite{Dress2002} to describe the band structure of ABA trilayer graphene. The Hamiltonian contains six tight-binding parameters that describes hopping between different sublattices. We denote  as $A_i$ ($B_i$) atoms from $A$ ($B$) sublattice, and index $i=1\ldots 3$ labels three layers.  Parameter $\gamma_0$ controls $A_i\leftrightarrow B_i$ hopping within the same layer; $\gamma_1$ determines the hopping between atoms atop of each other, $B_{1,3}\leftrightarrow A_{2}$ in our notations. Next, the parameter $\gamma_3$ corresponds to hops $A_{1,3}\leftrightarrow B_2$ and determines the trigonal warping. Parameter $\gamma_4$ labels hopping amplitude between atoms from same sublattices on adjacent layers, $A_{1,3}\leftrightarrow A_2$ and $B_{1,3}\leftrightarrow B_2$. Finally, much weaker parameters $\gamma_2$ and $\gamma_5$ determine hoppings between two outer layers, $A_1\leftrightarrow A_3$ and $B_1\leftrightarrow B_3$ respectively. 

In addition, we introduce the parameter $\delta$ to account for an extra on-site potential energy for $B_1$, $A_2$ and $B_3$ sites which are on top of each other. Parameters $\Delta_{1,2}$ are used to describe the effect of external electric field and charge asymmetry between middle and outer layers of the ABA-stacking graphene. They are related to the layer potentials $U_{1}\ldots U_{3}$ as:~\cite{Lu-mult,Guinea-mult,MinAbInitio,McCannGateABA}
\begin{equation}   \label{Eq:D1D2def}
\Delta_1  =  (-e)\frac{U_1-U_2}{2},
\quad
\Delta_2
=
(-e)\frac{U_1-2U_2+U_3}{6}.
\end{equation}
We note that the above parameterization is spin-independent: in the absence of the magnetic field, the spectrum is doubly spin-degenerate. 

The complete Hamiltonian can be separated into the trilayer $H_0$ in the absence of external electric field and $H_{\Delta_1}$,
\begin{equation}\label{Eqn:Hamiltonian}
H = H_0 + H_{\Delta_1}.
\end{equation}
We choose as our basis the atomic orbitals $A_1,B_1,A_2,B_2,A_3,B_3$ and write the two terms in the Hamiltonian as:
\begin{eqnarray}\label{Hamiltonian}
&&H_0=  \\
&&\begin{pmatrix} \Delta_2 & \gamma_0 t^*_k& \gamma_4t^*_k & \gamma_3 t_k & \frac{\gamma_2}{2}&0\\ 
\gamma_0 t_k & \delta + \Delta_2 & \gamma_1 & \gamma_4 t^*_k&0&\frac{\gamma_5}{2} \\
\gamma_4 t_k& \gamma_1 & \delta -2\Delta_1&\gamma_0 t^*_k&\gamma_4 t_k & \gamma_1 \\
\gamma_3t^*_k& \gamma_4 t_k & \gamma_0 t_k & -2\Delta_2 & \gamma_3 t^*_k& \gamma_4 t_k \\
\frac{\gamma_2}{2} & 0 & \gamma_4 t^*_k& \gamma_3 t_k & \Delta_2 & \gamma_0 t^*_k\\
0&\frac{\gamma_5}{2} & \gamma_1 & \gamma_4 t^*_k & \gamma_0 t_k & \delta + \Delta_2
\end{pmatrix},\\
&&H_{\Delta_1}  = \diag(\Delta_1,\Delta_1,0,0,-\Delta_1,-\Delta_1).
\end{eqnarray}

In Eq.~(\ref{Hamiltonian}), $t_k$ is a function of quasi-momentum $\bm{k}$,
\begin{equation}\label{QuasiMomentum}
t_k = \sum_{i=1}^3 \exp(i \bm{k}\cdot\bm{a}_i) = -1 -2 \exp(\sqrt{3}ik_y/2) \cos \frac{k_x}{2},
\end{equation}
where the summation is carried over position vectors $\bm{a}_i$ connecting an $A_1$ site to its nearest neighbors in a single honeycomb lattice: $\bm{a}_1 = (0,1/\sqrt{3})$, $\bm{a}_{2,3} = (\mp1/2,-1/2\sqrt{3})$. Quasi-momenta are given in units of inverse lattice constant $a=2.46 \,{\AA}$.

The values of tight-binding parameters are usually determined by matching the tight-binding structure to the experimental data.  Some of the parameter sets may be found in  Refs.~\onlinecite{Shimazaki2016,Campos2016,Datta2017,Datta2018}. In what follows we adopt the values of tight-binding parameters determined in Ref.~\onlinecite{Rao2018} using a combination of experimental data at zero magnetic field and Landau level spectrum. The values of these parameters read:
\begin{equation}\label{TBparameters}
\begin{aligned}
\gamma_0 = 3.1~\text{eV},~\gamma_1=0.38~\text{eV},~\gamma_2=-21~\text{meV},\\
\gamma_3 = 0.29~\text{eV},~ \gamma_4= -0.141~\text{eV},~\gamma_5= 50~\text{meV},\\ \delta = 35.5~\text{meV},~ \Delta_2 = 3.5~\text{meV}.
\end{aligned}
\end{equation}

At low energies, we expand $t_k$ from Eq.~(\ref{QuasiMomentum}) in quasi-momentum near its two minima $K^+$ and $K^-$, which are located at $(\pm4\pi/3,0)$ in the hexagonal Brillouin zone. Correspondingly in Eq.~(\ref{Hamiltonian}), we replace $\gamma_i t_k$ with $v_i\pi$ where
\begin{equation}
\pi = \xi k_x +i k_y, ~ \hbar v_i = \frac{\sqrt{3}}{2} a \gamma_i,
\end{equation}
with $\xi = \pm 1$ for $K^+$ and $K^-$ points respectively. 

In the absence of external electric field, the Hamiltonian (\ref{Hamiltonian}) can be shown via a change of basis to consist of monolayer- and bilayer-like bands.~\cite{McCannGateABA} The new basis is 
\begin{equation}
\bigg( \frac{A_1-A_3}{\sqrt{2}},\frac{B_1-B_3}{\sqrt{2}}, \frac{A_1+A_3}{\sqrt{2}}, B_2, A_2, \frac{B_1+B_3}{\sqrt{2}} \bigg)
\end{equation}
and the Hamiltonian now acquires the form 
\[
H = \begin{pmatrix}H_{\text{SLG}} & V_{\Delta_1} \\ V^T_{\Delta_1} & H_{\text{BLG}}\end{pmatrix},
\]
where blocks are defined in Eq.~(\ref{Hamiltonian1}).

\section{Calculation of Form Factors and The Exchange Integrals\label{Appendix:FormFactors}}
In this appendix, we obtain analytical expressions for the form factors and exchange integrals in Eq.~(\ref{ExchangeEnergy1}). The Hamiltonian has $N$ ($N=3$ in physical case) gullies and is $C_N$ symmetric. In the neighborhood of $i$-th gully center, the Hamiltonian without magnetic field has the form:
\begin{equation}\label{GullyHamiltonian}
H = \frac{p^{(i)2}_x}{2m_x} + \frac{p^{(i)2}_y}{2m_y}.
\end{equation}
Letting the 1st gully to have a center at $p_y=0$ (i.e.~on the $p_x$ axis), each $\bm{p}^{(i)}$ is given by successive rotation by an angle $\theta_i$ with $\theta_1=0$. The energy spectrum and eigenstates for (\ref{GullyHamiltonian}) in a magnetic field is found in Ref.~\onlinecite{Li2016}. Introducing the anisotropy parameters for the i-th gully $\alpha_i = \eta \cos{\theta_i} + i\sin{\theta_i}/\eta $, $\beta_i = \cos{\theta_i}/\eta + i\eta \sin{\theta_i}$, $\eta = (m_y/m_x)^{1/4}$ and the creation operator:
\[
\hat{a}^\dagger_i = \frac{1}{\sqrt{2}\hbar} (\alpha_i p_x + i \beta_i p_y),
\]
it is straightforward to verify that the Hamiltonian in each gully can be written as:
\[
H = \hbar \omega (\hat{a}^\dagger_i \hat{a}_i + \frac{1}{2}),
\]
where $\omega = eB/\sqrt{m_xm_y}$. We have used the Landau gauge $A_x=0, A_y= eBx$ and made the substitution $x\rightarrow x-X$ where $X$ is the electron orbital center. $l_B$ has been taken unity. The eigenstates and energy spectrum are found similarly to the case of a linear oscillator resulting in:
\begin{align}\label{AnisotropicWF}
\phi_{inX} (x,y) &= A_n e^{i \bm{Q}_i\cdot\bm{r}} e^{i X y-{\gamma^*_i x^2}/{2}} H_n \bigg(\frac{x}{|\alpha_i|}\bigg);&\\
E_n &= \hbar \omega \bigg(n+\frac{1}{2}\bigg).&
\end{align}
$A_n= (2^n n! \sqrt{\pi} |\alpha_i| L_y)^{-1/2}$ is the normalization factor, $\bm{Q}_i$ is the distance of gully to the K point. $\gamma_i = \beta_i/\alpha_i$, $i, n$ are gully and LL indices respectively. 

Substituting Eq.~(\ref{AnisotropicWF}) into (\ref{Eqn:Interactions}), one obtains matrix elements of the form $\langle i_1,n_1, X_1| \exp(i\bm{q}\cdot\bm{r}) | i_2,n_2, X_2\rangle$. The expression is evaluated by integrating over $y$ first, which produces a Kronecker delta $\delta_{X_1,X_2+q_y}$ and $x$ appear in the wave functions in the form $x-X_1$ or $x-X_2$. By a change of variable $z = x-X_1$, the expression gains a phase factor $\exp(i q_x X_1)$ and the integrand becomes independent of $X$. This allows a complete cancellation of all degeneracy indices after taking into account Kronecker delta-functions coming from expectation values of the form $\langle a^\dagger_{i_1n_1X_1}a^{}_{i_3n_3X_3}\rangle$ and summing over all $X$. This leads to Eq.~(\ref{ExchangeEnergy}) for the final expression for $U_{\text{ex}}$. Each form factor is given by $F^{nn}_{ik} (\bm{q}+\bm{Q}_{ki})$, where $\bm{Q}_{ik} = \bm{Q}_i-\bm{Q}_k$ is the momentum transfer between two gullies and $n$ is the gully LL index of the triplet. The analytical expressions for form factors calculated with the zeroth gully LL reads:
\begin{equation}\label{AnisotropicFormFactors}
F^{00}_{ik} (\bm{q}) = A'_{ik} \exp \bigg[iq_xq_y\bigg(\frac{1}{2}-\omega_{ik}\bigg)\bigg] \exp\bigg(-\frac{q_x^2 + \gamma_i\gamma^*_k q_y^2}{2 (\gamma_i +\gamma^*_k)} \bigg).
\end{equation}
$A'_{ik}=\surd(2/|\alpha_i||\alpha_k| (\gamma_i +\gamma^*_k))$, $\omega_{ik}=\gamma^*_k/(\gamma_i+\gamma^*_k)$. It is easy to see that inter-gully scattering is suppressed exponentially by the intergully distance $\bm{Q}_{ik}$. In the isotropic limit $\gamma_i = |\alpha| = 1$, Eq.~(\ref{AnisotropicFormFactors}) becomes the standard form factor obtained in Ref.~\onlinecite{MacDonald1984}.

\section{Proof of Gully Polarization theorem\label{Appendix:Proof}}
We prove the statement in Sec.~\ref{Sec:InteractionEffects} that, to first order in anisotropy and inter-gully scattering, the state that minimises (\ref{ExchangeEnergy1}) is either strongly gully polarized or fully gully coherent. As discussed in Sec.~\ref{Sec:InteractionEffects}, the exchange energy is reduced to Eq.~(\ref{ExchangeEnergy3}). This is to be minimized with the constraint $\sum |c_i|^2=1$. We choose the Lagrange multiplier to be $-2\lambda$ and solve for $\lambda$. Substituting the expression for $\lambda$ back into the equations gives for each $i$:
\begin{multline}\label{ExchangeMinimization}
(J_0-J_1) c_i \bigg(|c_i|^2 - \sum_k |c_k|^4\bigg) + \\J_2 \bigg(\sum_{m\ne i} c_m - c_i \sum_{k\ne m} c^*_k c_m\bigg) = 0.
\end{multline}
It is easy to see that setting $c_i$ as all real in Eq.~(\ref{ExchangeEnergy3}) results in a similar equation which does not affect the nature of the solution. In transforming the $J_1$ term, we use the identity $\sum_{k\ne i}|c_k|^2 = 1 - |c_i|^2$. Without inter-gully scattering, $J_2=0$, the first term has as a solution both, complete gully coherent state, $|c_i|=1/\sqrt{3}$, and gully polarized state $c_i =1, c_k = 0, i\ne k$. It is important that gullies are anisotropic, so that $J_0 > J_1$. The exchange energies calculated from the corresponding solutions are $-(J_0+2J_1)/6$ and $-J_0/2$ respectively, so the completely gully-polarized state is indeed the global minimum.  In the isotropic limit $J_0=J_1$, all choices of $c_i$ give the same energy corresponding to presence of full SU$(3)$ symmetry in the system.~\cite{Comment}

For a non-zero $J_2$ the bracketed expression proportional to $J_2$ admits the solution $c_1=c_2=c_3 = 1/\sqrt{3}$ modulus an arbitrary phase factor. Thus full gully coherence state is still an extrema. On the other hand, the completely gully-polarized state receives corrections and has components also in other gullies. Thus the non-zero $J_2$ removes the complete gully polarization.  A first order transition occurs when the energy for the gully-coherent state becomes a global minimum. By numerically minimizing Eq.~(\ref{ExchangeMinimization}), we find that this occurs when $J_2/(J_0-J_1)\sim0.25$.

\section{Details of the restricted Hartree-Fock calculation\label{Appendix:HartreeFock}}
In this Section we describe the Hartree-Fock (HF) approximation for completely filled Landau Levels (LL) originally proposed in Ref.~\onlinecite{MacDonald1984}. The essence of the method is a variational optimization of the energy over a trial set of wave functions (Slater determinants). In this work we largely follow the approach of Ref.~\onlinecite{Zhang2012}.  We aim to capture the interactions-induced splitting of emergent (nearly) three-fold degenerate Landau levels formed at large $\Delta_1$. In what follows we refer to such states as ``triplets'', where three-fold degeneracy originates from the set of three Dirac cones related to each other via $C_3$ rotation symmetry. Hence, we restrict our set of variational states to an arbitrary superpositions of single-particle triplet wave functions. 

\subsection{Numerical Procedure}
More specifically, we start with the set of six Landau level wave functions denoted as $\psi^{(ms)}_{\text{tri}}$, $m=1,2,3$. Index $s$ labels spin projection onto $z$-axis, so that $\psi^{(m\uparrow)}_{\text{tri}} =\psi^{(m)}_{\text{tri}}\otimes |\uparrow\rangle$ and $\psi^{(m,\downarrow)}_{\text{tri}} =\psi^{(m)}_{\text{tri}}\otimes |\downarrow\rangle$, with the wave function $\psi^{(m)}_{\text{tri}}$ obtained from exact diagonalization of Hamiltonian~(\ref{Eqn:Hamiltonian}). Three states $\psi^{(m)}_{\text{tri}}$ with $m=1,2,3$ can be distinguished by their transformation under $C_3$ rotations which can be intuitively seen as a proxy of ``angular momentum''. Due to presence of discrete rotational symmetry, this ``angular momentum'' is defined module 3 and takes values $0$, $1$, and $2$, corresponding to phase of $0$, $2\pi/3$ and $4\pi/3$ acquired from rotation by angle of $2\pi/3$. 

The wave functions $\psi^{(m)}_{\text{tri}}$ are vectors in the basis of Landau level indices and sublattices. They are obtained by exact diagonalization of the Hamiltonian near $K^{\pm}$ points; see Sec.~\ref{Sec:MagneticField}. Note, that the gully indices are omitted since all 3 Landau level forming the triplet belong to the same gully. In addition, we introduce a LL index cut-off $\Lambda_{\text{max}}=15$ which allows to represent triplet vector norm of more than $0.9$ in $\Delta_1$ range concerned, thus incorporating most of the tripltets weight. 

Projecting Hamiltonian on the manifold of 6 triplet states, we get the following expression:
\begin{multline}\label{HF1}
\langle m, s|H|m', s' \rangle = E_0(m)\delta_{m,m'}\delta_{s,s'}-E_{ZM} \sigma^z_{ss'} \\
+(U_H)^{ms}_{m's'}+ J^{ms}_{m's'}.
\end{multline}
In this Hamiltonian,  $E_0(m)$ represents the diagonal spin-degenerate single-particle Hamiltonian. The second term is the Zeemann energy which retains its standard form after projection onto the triplet states. The last two terms in Eq.~(\ref{HF1}) originated from the interactions and account for Hartree and exchange terms respectively. These terms can be obtained from the transformation of conventional  Hartree and exchange terms by the wave functions of triplet states. Thus these terms depend on the density matrix in the basis of sublattices ($\alpha,\alpha'$) and  Landau levels ($n,n'$), $\Delta^{\alpha' n' s'}_{\alpha n s}$. This density matrix can be obtained from the density matrix in the triplet basis, $\Delta^{m_is_i}_{m_ks_k}$ via the change of basis:
\begin{equation}\label{DM}
\Delta^{\beta n' s'}_{\alpha n s} = \sum_{m_i,m_k,s_i,s_k} \Delta^{m_is_i}_{m_ks_k}\psi^{(m_is_i)}_{\alpha n s} \otimes \psi^{(m_ks_k)\dagger}_{\beta n' s'}.
\end{equation}

Using density matrix in the basis of Landau levels, $\Delta^{\beta n' s'}_{\alpha n s}$, we can write standard expressions for Hartree and exchange terms, following Ref.~\onlinecite{Zhang2012}:
\begin{equation}\label{ABAHamiltonian}
\begin{aligned}
\langle \alpha n s |U_{H}|\beta n' s'\rangle &= \frac{E_H}{2}\Delta_\text{mid} (2\delta_{B_2,\alpha}+2\delta_{A_2,\alpha}-1),&
\\ 
\langle \alpha n s |U_{ex}|\beta n' s'\rangle &=J^{\alpha\beta s s'}_{n,n_1,n_2,n'}\Delta^{\beta n_2 s'}_{\alpha n_1 s}.&
\end{aligned}
\end{equation}
where parameter $E_H={e^2 d}/({2 l^2_B \kappa_0})$ characterizes the scale of the Hartree energy. Here $e$ is the electron charge, $\kappa_0$ is the effective screening constant and $d=0.335$~nm measures the distance between adjacent graphene layers. Density matrix projection $\Delta_\text{mid} = \sum_{n,s} (\Delta^{A_2 n s}_{A_2 n s}+\Delta^{B_2 n s}_{B_2 n s})$ corresponds to the electron density on the middle layer. In this paper, we assume $U_H$ has been neutralized by a positive charge background and set it to zero. 

The exchange integral is defined as: 
\begin{equation}
J^{\alpha\beta s s'}_{n,n_1,n_2,n'} = \int \frac{d^2q}{(2\pi)^2} U_{\alpha \beta}(q) F_{n,n_1}(-q) F_{n_2,n'}(q)  \delta_{ss'}.
\end{equation}
The explicit form of the form factors $F_{nn'}(q)$ is listed in Ref.~\onlinecite{MacDonald1984}, and the interaction potential in the exchange integral is given by:
\begin{equation}\label{Potential}
U_{\alpha\beta}(q) =  \frac{2\pi e}{q \varepsilon(q)} T_{\alpha \beta}
\end{equation}
where $\varepsilon(q)$ is the dielectric function. $T_{\alpha \beta} = 1, \exp (-qd)$ or $\exp(-2qd)$ for $\alpha,\beta$ in the same, adjacent or  different outer layers.

The projection of the exchange interaction matrix onto the triplet basis is given by:
\begin{equation}
J^{m_is_i}_{m_ks_k} = \sum \psi^{(m_ks_k)}_{\beta n' s'} \langle \alpha, n, s |U_{ex}|\beta, n', s'\rangle  \psi^{(m_is_i)\dagger}_{\alpha n s},
\end{equation}
where the summation is taken over repeated indices. 

The self-consistent solution of HF equations is implemented  as follows. For instance, fixing filling at $N=1$, we start with the trial density matrix in the triplet basis, $\Delta^{m_is_i}_{m_ks_k} = (c_1,c_2,c_3)\times (c_1,c_2,c_3)^\dagger |\uparrow\rangle \langle\uparrow|$, where $c_i$ are random normalized coefficients $\sum_{i=1}^3 |c_i|^2=1$. Using this density matrix, we calculate the density matrix in LL basis and exchange integrals according to Eqs.~(\ref{DM})-(\ref{Potential}). Finally, by diagonalizing projected Hamiltonian in  Eq.~(\ref{HF1}) we calculate updated eigenstates $|n\rangle$ and produce a new density matrix $\Delta^{m_is_i}_{m_ks_k}$ by filling the lowest $\nu$ of them ($\nu$ is fixed to $\nu=1$ in what follows), 
\[
\Delta^{m_is_i}_{m_ks_k} = \sum_{n=1}^\nu |n \rangle \langle n|. 
\]
For probability density plots in the main text, the above procedure is repeated until eigenvalues and eigenstate coefficients converge. For dipole moment plots, iteration number is set to be 500.

Intuitively, one can easily undertand why the interactions favor the symmetry broken state at $\nu=1$. Each of the single-particle wave functions $\psi^{(m)}_\text{tri}$, $m=1,2,3$ lives on all three Dirac points (see Fig.~\ref{Fig:LLplots} in the main text). In fact, in the limit of weak magnetic field (or large separation between emergent Dirac gullies), these single particle wave-functions become the proper combination of wave-functions localized on each of the Dirac cones $\phi_i$ with an additional phase factors
\begin{eqnarray}\label{Eqn:C3Superposition}
\psi^{(1)}_\text{tri} &=& \frac{1}{\sqrt 3} (\phi_1+\phi_2+\phi_3),\\
\psi^{(2)}_\text{tri} &=& \frac{1}{\sqrt 3} (\phi_1+e^{2\pi i/3}\phi_2+e^{4\pi i/3}\phi_3),\\
\psi^{(3)}_\text{tri} &=& \frac{1}{\sqrt 3} (\phi_1+e^{4\pi i/3}\phi_2+e^{2\pi i/3}\phi_3).
\end{eqnarray}
The $C_3$ rotations simply permutes $\phi_i$ between themselves. This results in the function $\psi^{(1)}_\text{tri}$ being invariant under rotation, and remaining two states $\psi^{(2,3)}_\text{tri} $ acquiring a phase factor $e^{\pm 2\pi i/3}$. Now, since support of wave functions $\phi_i$ and $\phi_j$ are weakly overlapping for $i\neq j$, exchanges favor the state where all weight of the wave function is located in one of the Dirac gullies. In the basis of $\psi^{(m)}_\text{tri}$ such state corresponds to a coherent superposition of all three single-particle wave functions and it breaks $C_3$ rotation symmetry.

\subsection{Screening}\label{Appendix:Screening}
In the reduced basis of one triplet, Coulomb interactions between three LLs receive polarization corrections from all other LLs. As a result, the dielectric function $\varepsilon(q)$ acquires a non-trivial dependence on $q$.  Asymptotic behavior of $\varepsilon(q)$ was derived in the large and small-$q$ limit in Ref.~\onlinecite{Gorbar2010}. For HF calculations in this paper, we use an interpolation formula proposed in Ref.~\onlinecite{Papic2014} for $\varepsilon(q)$:
\begin{equation}\label{DielectricFunction}
\varepsilon(q) = 1 +  \frac{f({q^2l^2_B}/{2})}{ql_B},
\end{equation}
where function $f(x)=a \tanh(1.25 x)$ and parameter $a \propto me^2/\kappa_0\hbar^2$ is a dimensionless constant whose value depends on the specific system. Quasi-particle mass is set to $m=\sqrt{m_xm_y}\sim 0.005 m_e$ for the $\Delta_1$ range considered for HF calculations, where $m_e$ is the electron mass. Given the overall good agreement of experimental data with single-particle simulations in Ref.~\onlinecite{Rao2018}, we expect that the LL mixing and interaction corrections must be smaller than typical cyclotron gaps. In this paper we choose $a=10$ which gives, for example, $J_{11,11}\sim3$~meV.

\subsection{Visualizing symmetry broken states}\label{Appendix:QuasiClassicalWF}

In order to visualize the form of the symmetry broken states in real space, we transform the LL wave functions into the maximally localized ``wave packet''. This is done via convolving the single particle LL wave function in the Landau gauge with the Gaussian envelope function, 
\[
\Psi_{n} (x,y) = \int_{-\infty}^{\infty} C_X \exp(i X y / l_B^2) \psi_n\bigg(\frac{x-X}{l_B}\bigg) dX
\]
where $\psi_n$ is the $n$-th eigenstate of the Hamiltonian. In order to get the maximally localized wave packet in both directions, we choose $C_X = (2\pi l_B^2)^{-\frac{1}{2}} \exp(-X^2/2l_B^2)$. We calculate the integral using explicit expression for $\psi_n$,
\[
\psi_n(x) = \frac{1}{\pi^{\frac{1}{4}} \sqrt{2^n n! l_B}} \exp(- x^2 /2l_B^2) H_n(x),
\] 
where $H_n(x)$ is the $n$-th Hermite polynomial. This gives the following wave function describing LL ``wave packet'' centered at the origin:
\begin{multline}\label{Eqn:QuasiClassicalWF}
\Psi_{n} (x,y) = \frac{1}{\sqrt{n!}} \bigg(\frac{x-iy}{\sqrt{2}l_B}\bigg)^n  \exp \bigg(-\frac{x^2+y^2}{4 l_B^2}+i \frac{xy}{2 l_B^2}\bigg).
\end{multline}

We numerically simulate the probability distribution for the triplet eigenstates $\psi^{(m)}_\text{tri}$, $m=1,2,3$ at $B=1.25$~T and compare them with the momentum band structure. More specifically, we plot probability density $p(x,y)$ for the wave function in the basis of LL and sublattices, $\psi^{\alpha n}$, which is calculated as 
\begin{equation}
p(x,y)
=
\sum_{\alpha=1}^6\left|
\sum_{n=1}^{\Lambda_\text{max}} c_{\alpha n} \Psi_{n} (x,y)
\right|^2,
\end{equation}
where the inner sum goes over LL and outer sum sums probability density for each of the sublattices.
\begin{figure}[h]
	\includegraphics[width=\linewidth]{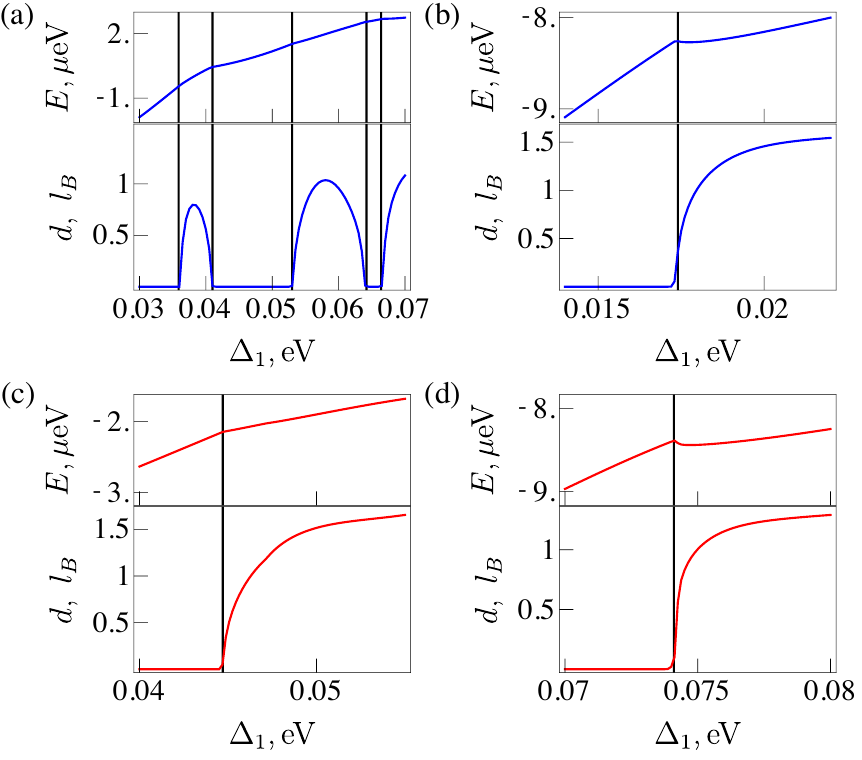}
	\caption{\label{Fig:EnergyDipole}  Panels (a)-(d) show dipole moments and HF energies as a function of $\Delta_1$ for T1-4 respectively. At critical points marked by vertical lines, the HF energies demonstrate cusps characteristic of a first order transition.}
\end{figure}

\subsection{Dipole moment calculation}\label{Appendix:DipoleMoment}
The existence of the first order phase transition can be verified by measuring the dipole moment $\bm{d}$ of the system. Expectation value of dipole moment, $\bm{d}$, is found by averaging the position of an electron over the HF ground state. This is most easily done in the quasi-momentum representation where the problem of degeneracies does not arise. The $n$-th LL eigenstate can be shown to be:
\begin{equation}\nonumber
\Psi_{\alpha n}(\bm{k}) \propto \exp(-ik_xk_y)\psi_{\alpha n} (k_x),
\end{equation}
where $\alpha$ denotes layer and sublattice indices and $\psi_{\alpha n} (k_x)$ is the $n$-th momentum eigenstate of a harmonic oscillator at $\alpha$. Let $a^\alpha_n$ be the corresponding component of the HF state in the global basis, simple calculations show that
\begin{equation}
\bar{x} = \sum_{m,n,\alpha} x_{mn} a^{\alpha*}_ma^\alpha_n; ~ \bar{y}=\sum_{m,n,\alpha} p_{mn} a^{\alpha*}_m a^\alpha_n.
\end{equation}
$x_{mn}$ and $p_{mn}$ are coordinate and momentum matrix elements of a harmonic oscillator. The averaged position vector $\bar{\bm{r}}$ seems to agree qualitatively with our visual representation of the probability density: for a given state, $\bar{\bm{r}}$ is approximately the sum of position vectors of each gully $\bm{r}_i$ weighted by their respective probability: $\bar{\bm{r}} = \sum_i |c_i|^2\bm{r}_i$. In particular, $\bar{\bm{r}}$ vanishes for the $C_3$ symmetric single particle LLs due to the mod 3 feature described in Sec.~\ref{Sec:MagneticField}.

Finally, we offer additional evidence in support of the conclusion that dipole transitions in Sec \ref{SubSec:Numerics} is first order. In Fig.~\ref{Fig:EnergyDipole}, we plot and compare dipole moments and HF energies as a function of $\Delta_1$ for T1-4. It shows that critical points for dipole moments coincide with a cusp in the HF energies, indicating a first order transition in which one minima overtakes the other.

%\bibliography{refs2}

\begin{thebibliography}{40}
\bibitem [{Com({\natexlab{b}})}]{Comment}%
\BibitemOpen
\href@noop {} {}\bibinfo {note}  {Isotropy in fact is not
		essential for $J_0 = J_1$. As can be seen from Eq.~(17), the equality would
		also hold for exchange between two anisotropic gullies with the same
		orientation that are displaced with respect to each other in momentum
		space.}\BibitemShut {Stop}%
\end{thebibliography}

\end{document}